\documentstyle[12pt]{article}
\displaywidth\textwidth
\linewidth\textwidth
\begin{document}
\title{Defect--Defect Correlation Functions, Generic Scale
Invariance, and the Complex Ginzburg--Landau Equation}
\author{Bruce W. Roberts, Eberhard Bodenschatz, and James P. Sethna}
\date{Laboratory of Atomic and Solid--State Physics,\\
Cornell University,\\
Ithaca,  NY 14853--2501.}
\maketitle


\bigskip


\begin{abstract}
Motivated by the idea of developing a ``hydrodynamic'' description
of spatiotemporal chaos, we have investigated the defect--defect
correlation functions in the defect turbulence regime of the
two--dimensional, anisotropic complex Ginzburg--Landau equation.
We compare our results with the predictions of generic scale
invariance.  Using the topological nature of the
defects, we prove that defect--defect correlations
cannot decay as slowly as predicted by generic scale invariance.
We also present results on the fluctuations of the
amplitude field $A$.
\end{abstract}

\newpage

\section{Introduction}
\label{CGLsec:introduction}

Chaos is the name given to intrinsic random behavior arising in a
deterministic system \cite{CGLref:Ott}.  The simplest forms of chaos
occur in systems of three or more coupled ordinary differential
equations and in discrete mappings.  A great deal is known about
such systems with a small number of degrees of freedom displaying
``temporal chaos'', where the structure of the phase space can be
analyzed in detail.  Spatially extended systems with many
interacting degrees of freedom exhibit
``spatiotemporal chaos'' \cite{CGLref:CrossHo}, in which the chaotic
fluctuations occur on spatial scales as well as in time.
Spatiotemporal chaos has also been called
``weak turbulence'', ``defect chaos'', and ``defect turbulence''.
(This is to be distinguished from fully developed turbulence, which
typically occurs for increasing driving but fixed size $L$, while
spatiotemporal chaos occurs for fixed large driving but
increasing $L$.)
There have been a number of approaches used in attempting to understand
spatiotemporal chaos: partial differential equations, coupled ordinary
differential equations (in which space is discretized),
coupled map lattices \cite{CGLref:Kaneko}
(in which space and time are discretized),
and cellular
automata (in which space, time, and dependent variable are discretized).
For large systems,
we may hope to use a statistical description of the chaotic
states, borrowing concepts from equilibrium statistical mechanics.  We
wish to find simple reduced descriptions, emphasizing the
collective behavior of many chaotic degrees of freedom, analogous to
the reduced long--wavelength description provided by thermodynamics
and hydrodynamics for equilibrium systems
\cite{CGLref:ciliberto,CGLref:millerhuse,CGLref:hohshrai,CGLref:cmlchaos,CGLref:hwa,CGLref:crosstu}.

Typically, spatiotemporal
chaos appears in nonequilibrium pattern--forming systems slight\-ly
above their threshold of instability \cite{CGLref:CrossHoRev}.  The
study of spatiotemporal chaos has been advanced by the development of
experimental systems which are precisely controlled and have a large
aspect ratio
\cite{CGLref:rbconvectionI,CGLref:rbconvectionII,CGLref:canada,CGLref:Steinberg,CGLref:ehcI,CGLref:ehcII,CGLref:ehcIII,CGLref:capillary,CGLref:NonlinearOptics}.  Such
systems have large statistically homogeneous regions relatively free
from boundary effects.  A key question to address is whether such
regions can be described in terms of hydrodynamic--like theories,
focusing on collective behaviors and long--wavelength descriptions.
We wish to address an aspect of this question by considering the
coherent structures known as topological defects (or spirals or
vortices) in a system that exhibits spatiotemporal chaos: the complex
Ginzburg--Landau equation
\cite{CGLref:defturbI,CGLref:defturbII,CGLref:defturbIII,CGLref:defturbIV,CGLref:aransoninteractI}.
This equation describes the slowly varying amplitude and phase in an
extended system which undergoes a supercritical Hopf bifurcation to an
oscillating and spatially uniform or oscillatory and spatially
periodic state.  The equation exhibits many interesting patterns, but
we will restrict our investigation to the Benjamin--Feir unstable (or
defect turbulent) regime \cite{CGLref:Newell}, in which topological
defects occur in the context of spatiotemporal chaos.  Other systems,
such as Rayleigh--B\'enard convection
\cite{CGLref:rbconvectionI,CGLref:rbconvectionII,CGLref:canada},
electrohydrodynamic convection in liquid crystals
\cite{CGLref:ehcI,CGLref:ehcII,CGLref:ehcIII,CGLref:nasuno,CGLref:sasaI},
capillary ripples \cite{CGLref:capillary}, cardiac
tissue \cite{CGLref:heart}, chemical reactions \cite{CGLref:swinney},
and wide aperture lasers \cite{CGLref:NonlinearOptics,CGLref:LegaLaser},
can exhibit similar defect--turbulent behavior.  Here, we will examine
the defect--defect correlation functions and relate them to the ideas
of generic scale invariance, which is a theory for describing
nonequilibrium systems with conservation laws.
We will prove that the
generic predictions cannot be correct for topological defect
correlations.  Finally, we will see that our numerical
results do not agree with
the predictions of generic scale invariance
\cite{CGLref:gsiI,CGLref:gsiII,CGLref:gsiIII,CGLref:grinsand,CGLref:garrido}.

This paper is organized as follows.  In Section \ref{CGLsec:cgleqn}
we discuss briefly the complex Ginzburg--Landau equation.
The numerical technique that we use to solve this equation
is discussed in Section \ref{CGLsec:nummethod}.  Some
interesting results that don't depend solely upon the defects in the
system are presented in Section \ref{CGLsec:Aresults}.  In
Section \ref{CGLsec:gsi} we briefly describe the ideas behind
generic scale invariance, and relate these ideas to the topological
defects in the complex Ginzburg--Landau equation.
Constraints on the behavior of the defects due to their topological
nature are discussed in Section \ref{CGLsec:constraints}.
The predictions of generic scale invariance, as well as that from
the constraints, are compared with numerical results in
Section \ref{CGLsec:numerics}.  Finally, we present our
conclusions in Section \ref{CGLsec:conclusions}.

\section{The Complex Ginzburg--Landau Equation}
\label{CGLsec:cgleqn}

Perturbative analyses of microscopic equations for various pattern--forming
systems yield complex partial differential equations which go under
the name ``amplitude equations''.
Considered in their own right as model dynamical systems which do
not necessarily describe any real physical system, these equations are
referred to as Ginzburg--Landau models.  It is important to note the
role of these Ginzburg--Landau equations as model equations.  Many
properties of nonequilibrium systems are encountered in these
equations.  Many hard problems, such as the existence and interaction
of defects and coherent structures, or the appearance of chaos,
can be profitably addressed in the simple framework provided by
these equations.
However, these equations provide a
quantitative description of real experiments valid only in a small
region near the transition threshold for a pattern.
Far from threshold, only the phase of the complex field survives as
a slow degree of freedom, since it describes a symmetry of the system.
The magnitude of the complex field only becomes slow near threshold;
far away it is just one of the many fast degrees of freedom.

We wish to consider spatiotemporal chaos and topological defects
in an equation that has been used both as an ``amplitude equation''
in the sense discussed above as well as a model equation to study
generic features of spatially extended nonlinear dynamical systems.
We are using the equation as a model equation for spatiotemporal chaos;
its applicability to real physical systems is limited to the weakly
nonlinear regime.
The complex Ginzburg--Landau equation we consider is given by
\begin{equation}
\partial_t A =
	(\mu_1 + {\rm i}\mu_2) A - (c_1 + {\rm i} c_2) \vert A \vert^2 A +
	( b_{1x} + {\rm i} b_{2x} ) \partial_x^2 A +
	( b_{1y} + {\rm i} b_{2y} ) \partial_y^2 A
\label{CGLeqn:fullcgl}
\end{equation}
where $A$ is a complex field in two dimensions
(see for example \cite{CGLref:cglreview}).
By rescaling
$A' = \sqrt{\frac{c_1}{\mu_1}} A {\rm e}^{-{\rm i} \mu_2 t}$,
$t' = \mu_1 t$,
$x' = \sqrt{\frac{\mu_1}{b_{1x}}} x$,
$y' = \sqrt{\frac{\mu_1}{b_{1y}}} y$,
$b_x = \frac{b_{2x}}{b_{1x}}$,
$b_y = \frac{b_{2y}}{b_{1y}}$, and
$c = \frac{c_2}{c_1}$,
we obtain the following equation (dropping now the primes):
\begin{equation}
\partial_t A = A - (1 + {\rm i} c) \vert A \vert^2 A +
	( 1 + {\rm i} b_x ) \partial_x^2 A +
	( 1 + {\rm i} b_y ) \partial_y^2 A.
\label{CGLeqn:cgl}
\end{equation}
This equation
can have topological defect solutions where $A = 0$ (both ${\rm Re}[A]$ and
${\rm Im}[A]$ are zero)
\cite{CGLref:defturbI,CGLref:defturbII,CGLref:defturbIII,CGLref:defturbIV,CGLref:kuramoto,CGLref:defectsolutionsI,CGLref:defectsolutionsII,CGLref:huber,CGLref:sakaguchi}.
These defects can occur either in static arrangements or in
dynamic ones.  The dynamics states can have defects moving in a regular
fashion (for example, all defects drifting in one direction)
or they can exhibit
defect turbulence, where defects are
continuously nucleated and annihilated in pairs
and are moving about in a chaotic, non-regular fashion.
We wish to focus on the latter case,
the Benjamin--Feir turbulent instability regime
\cite{CGLref:Newell,CGLref:benjaminfeir,CGLref:eckbf,CGLref:bfanis},
which occurs when $1 + b_\alpha c < 0$.
In this region of parameter space, all spatially periodic solutions of the
complex Ginzburg--Landau equation are unstable.
For comparison with the
ideas of generic scale invariance
\cite{CGLref:gsiI,CGLref:gsiII,CGLref:gsiIII,CGLref:grinsand,CGLref:garrido},
we will focus on the anisotropic case $b_x \neq b_y$.

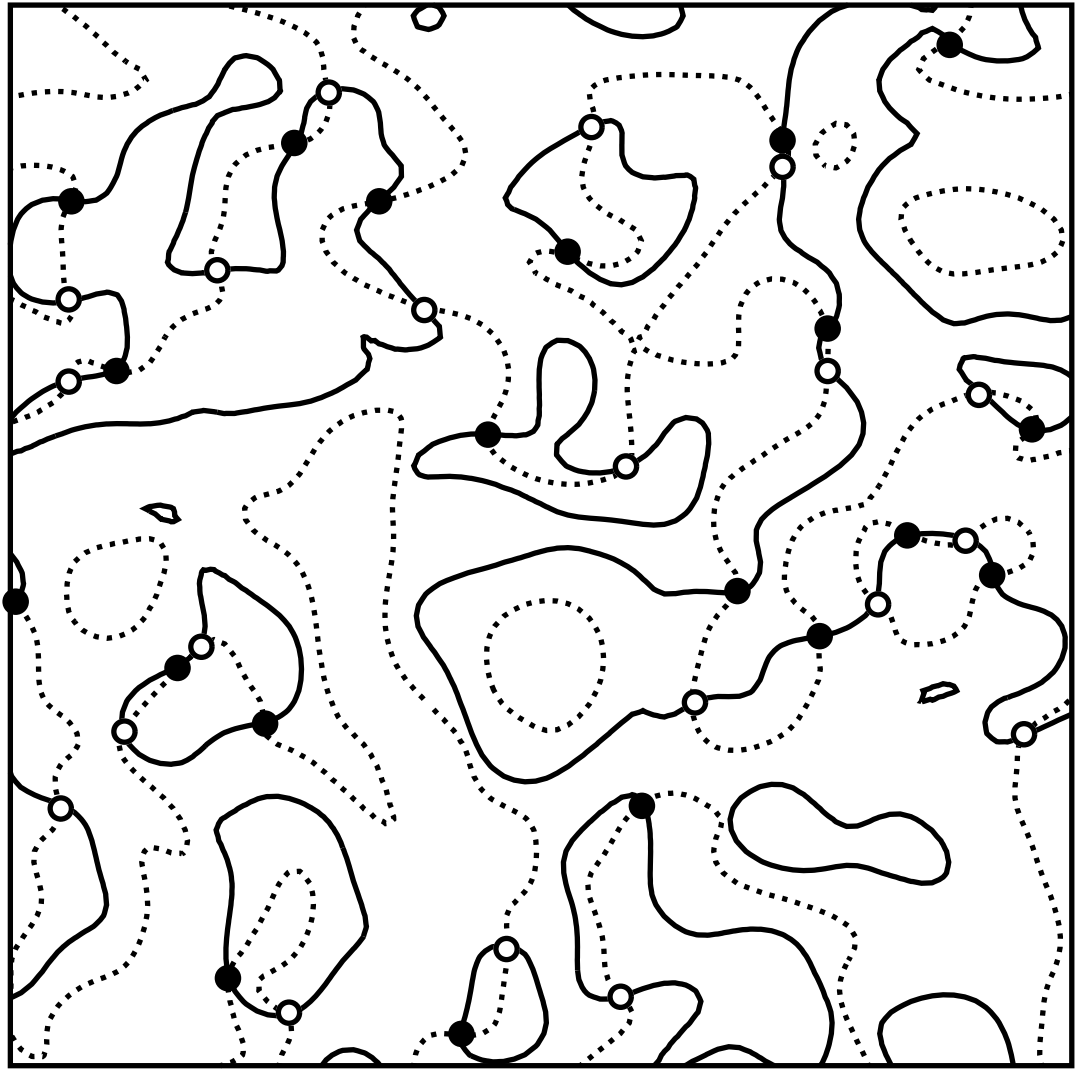
\begin{figure}
\caption[Snapshot of a $70 \times 70$ region]
{Snapshot of a $70 \times 70$ region. This is $\sim 1/12$ of the
total area of the simulation.  The solid
lines are where ${\rm Re}~[A]=0$, and the dashed lines
are where ${\rm Im}~[A] = 0$.
Filled circles ($\bullet$) are vortices with
topological charge $+1$, and the open
circles ($\circ$) have charge $-1$. The typical distance between
defects is 10.
This picture is for $c = 1.5$, $b_x = -0.75$, and $b_y = -3.0$.
}
\label{CGLfig:snapshot}
\end{figure}

The topological defects come in two varieties.  The type
of defect depends on how the phase of $A$ changes as we travel
counterclockwise once around the defect.  A defect with a phase jump
of $2\pi$ has a topological charge of $+1$, while one with a jump of
$-2\pi$ has a topological charge of $-1$.  This is analogous to the
right--handed and left--handed single--armed spirals in
Belousov--Zhabotinskii reactions \cite{CGLref:BZreact}.
Let $\rho_+({\rm \bf r})$ equal the
density of $+1$ defects and $\rho_-({\rm \bf r})$ equal the density of
$-1$ defects.  We can then define a ``topological'' order parameter,
$\rho({\rm \bf r}) \equiv \rho_+({\rm \bf r}) - \rho_-({\rm \bf r})$,
which is just the density of the defects weighted by their topological
charge.  This order parameter is conserved in a system with periodic
boundary conditions: $\int_V \rho({\rm \bf r}) {\rm d\bf r} = 0$, as
defects can only be created or destroyed in $+/-$ pairs.  We focus on
the order parameter $\rho({\rm \bf r})$ as an effective
coarse--grained field, which, for the purpose of testing the
applicability of generic scale invariance, we conjecture can be
described by some hydrodynamic equation of motion.
Figure \ref{CGLfig:snapshot} is a snapshot
of the simulated system, showing the two types of defects, and
lines of ${\rm Re}~[A] = 0$ and ${\rm Im}~[A] = 0$.

Although the time evolution of ensembles of defects is very complicated,
considerable progress has been made in the study of the dynamics of
isolated or weakly interacting defects.  Defects are isolated singularities
in the phase equation (for the phase of $A$), but are smooth solutions
of the full equations.  It is an attractive idea to imagine a description
in terms of coupled dynamics of phase and defect degrees of freedom.
Progress on this idea has been made in the case of
Rayleigh--B\'enard convection
\cite{CGLref:crosstu,CGLref:CrossNewell,CGLref:NPS}
and
the Kuramoto--Sivaskinsky equation \cite{CGLref:hwa};
for a review of the situation in the
complex Ginzburg--Landau equation see
\cite[pages 920--922]{CGLref:CrossHoRev}.

The complex Ginzburg--Landau equation has also been studied in one dimension,
where it also exhibits spatiotemporal chaos.
Much effort has been made in this case to apply ideas from low--dimensional
chaos, such as Lyapunov exponents and dimension densities
\cite{CGLref:hohshrai,CGLref:Shraietal,CGLref:Chate,CGLref:egolfnature,CGLref:bohrnature,CGLref:egolfprl}.
It is not clear that such ideas will be applicable to spatiotemporal
chaos in general, or to the two--dimensional complex Ginzburg--Landau
equation with topological defects in particular.
In one--dimension, an important question, still unsettled, is whether phase
turbulence (turbulence without strong amplitude fluctuations)
exists \cite{CGLref:egolfprl}.  In contrast, even an apparent phase turbulence
regime has not been seen in
the two--dimensional complex Ginzbug--Landau equation.  The two
dimensions (one versus two--dimensional) may very
well be fundamentally different.  Numerical simulations can provide
insight into the relationship between these regimes.

\section{Numerical Methods}
\label{CGLsec:nummethod}

Equation (\ref{CGLeqn:cgl}) is a continuum equation.  We wish to put
it in a form that allows approximate solution on a computer, which
requires a discretization scheme, an eigenfunction expansion scheme,
or some combination of both.  We use a technique that combines
elements of both approaches; it is a spectral method (for a general
review of this see \cite{CGLref:canuto,CGLref:orszag}).  This is a
special form of a more general technique known as the method of
weighted residuals \cite{CGLref:fletcherI,CGLref:fletcherII}.

Our original code was obtained from P. Coullet, L. Gil, and
J. Lega (see references
\cite{CGLref:defturbI,CGLref:defturbII,CGLref:defturbIII,CGLref:defturbIV,CGLref:coulletinit}),
and is now used as the basis for simulations by several groups.
The original code was simply the solution to the differential equation.
We have optimized the code, speeding it up by a factor
of $3$; furthermore, we have added sections for analysis of defects
and Fourier space quantities.

To begin our discussion of the numerical solution algorithm,
we rewrite equation (\ref{CGLeqn:cgl}) as
\begin{equation}
\partial_t A = A + B_x \partial^2_x A + B_y \partial^2_y A -
	C |A|^2 A
\label{CGLeqn:newcgl}
\end{equation}
where $B_x = 1 + {\rm i} b_x$, $B_y = 1 + {\rm i} b_y$,
and $C = 1 + {\rm i} c$.
Now, if we knew the exact solution, equation (\ref{CGLeqn:newcgl}) would
be satisfied everywhere in space.  However, we can only use a finite
number of basis functions or grid points in a numerical solution method,
so the solution we obtain is necessarily approximate.  In order to
make this approximation systematic, we define the residual
\begin{equation}
R({\bf x},t) \equiv \partial_t A - A + C \vert A \vert^2 A -
	B_x \partial_x^2 A - B_y \partial_y^2 A
\label{CGLeqn:residualdef}
\end{equation}
for the numerical solution. In order for our numerical solution to
be an accurate approximation to the full solution, this
residual needs to be small.  There are a number techniques
that accomplish this and also fall under the general
name ``method of weighted residuals''
\cite{CGLref:fletcherI,CGLref:fletcherII}.

To formulate the numerical solution, the field $A({\bf x},t)$
is expanded
\begin{equation}
A({\bf x},t) = \sum_{\alpha=1}^{N_B} A_\alpha(t) \phi_\alpha({\bf x})
\label{CGLeqn:expandA}
\end{equation}
where $\phi_\alpha({\bf x})$ is some basis function.  Then we demand that
\begin{equation}
\int R({\bf x},t) \psi_\beta({\bf x}) {\rm d \bf x} = 0,
\ \ \ \ \ \ \ \ \ \ \beta = 1,...,N_B
\label{CGLeqn:residualint}
\end{equation}
where $\psi_\beta({\bf x})$ is our weighting function.  A variety of
numerical techniques follow this general procedure.
For example, if
the basis set $\{ \phi_\alpha \}$ is the same as
$\{ \psi_\beta \}$ then the method
is known as a Galerkin technique.  The finite element method
can also be written
in this general form \cite{CGLref:fletcherI,CGLref:fletcherII}.

We use the ``Fourier collocation''
method \cite{CGLref:canuto,CGLref:orszag}.
Note that we are always using periodic boundary conditions.
For other types of boundaries (such as $A=0$ at the boundaries),
we could use Fourier sine transforms and expand in a basis set
of sine waves. Our approach is to use
\begin{equation}
\phi_\alpha({\bf x}) = {\rm e}^{{\rm i} {\bf k}_\alpha \cdot {\bf x}}
\label{CGLeqn:basis}
\end{equation}
where ${\bf k}_\alpha = ( \frac{2 \pi}{L_x} k_1, \frac{2 \pi}{L_y} k_2 )
= (k_x, k_y)$ for $k_1 = 0,...,N_x-1$ and $k_2 = 0,...,N_y-1$.
This is the ``Fourier'' part.  For the ``collocation'', the
residual is forced to be zero at the lattice points by choosing
a set of delta functions for the weight functions
\begin{equation}
\psi_\beta({\bf x}) = \delta({\bf x} - {\bf x}_\beta) =
	\delta(x - x_{j_1}) \delta(y-y_{j_2})
\label{CGLeqn:weights}
\end{equation}
where $x_{j_1} = j_1 \delta x = j_1 \frac{L_x}{N_x}$ and
$y_{j_2} = j_2 \delta y = j_2 \frac{L_y}{N_y}$, with $j_1 = 0,...,N_x-1$
and $j_2 = 0,...,N_y-1$.

At this point we define our conventions for the Fourier
transform.  The forward transform is
\begin{equation}
h_{k_1 k_2} = \frac{1}{N_x N_y} \sum_{j_1=0}^{N_x-1}
	\sum_{j_2=0}^{N_y-1} g_{j_1 j_2} {\rm e}^{- 2 \pi {\rm i}
	\frac{j_1 k_1}{N_x}}
	{\rm e}^{- 2 \pi {\rm i} \frac{j_2 k_2}{N_y}}
\label{CGLeqn:ft}
\end{equation}
and the inverse transform is
\begin{equation}
g_{j_1 j_2} = \sum_{k_1=0}^{N_x-1}
	\sum_{k_2=0}^{N_y-1} h_{k_1 k_2} {\rm e}^{ 2 \pi {\rm i}
	 \frac{j_1 k_1}{N_x}}
	{\rm e}^{ 2 \pi {\rm i} \frac{j_2 k_2}{N_y}}.
\label{CGLeqn:ftinv}
\end{equation}
We calculate the Fourier transforms in the algorithm using
IBM's Engineering and Scientific Subroutine Library (ESSL)
Fast Fourier Transform (FFT) \cite{CGLref:essl}.

We now want to substitute equations (\ref{CGLeqn:basis}) and
(\ref{CGLeqn:weights}) into equation (\ref{CGLeqn:residualint}).
We will show this for the term $\partial_x^2 A$.
Start with
\begin{equation}
\int \partial_x^2 A({\bf x},t) {\rm d \bf x} = - \sum_{k_1 k_2}
A_{k_1 k_2}(t) \left(\frac{2 \pi k_1}{L_x}\right)^2
	{\rm e}^{2 \pi {\rm i} \frac{k_1 x}{L_x}}
	{\rm e}^{2 \pi {\rm i} \frac{k_2 y}{L_y}}
	\delta(x-j_1 \delta x) \delta(y - j_2 \delta y) {\rm d}x {\rm d}y.
\label{CGLeqn:partial}
\end{equation}
Performing the integral and recalling the definitions of $\delta x$,
$\delta y$, and $k_x$ gives
\begin{equation}
\sum_{k_1 k_2} A_{k_1 k_2}(t) k_x^2
	{\rm e}^{2 \pi {\rm i} \frac{k_1 j_1}{N_x}}
	{\rm e}^{2 \pi {\rm i} \frac{k_2 j_2}{N_y}}.
\label{CGLeqn:partialint}
\end{equation}
Doing similar integrals for the other terms in $R({\bf x},t)$ and
then using the inverse Fourier transform on the expressions
(which amounts to dropping the terms
${\rm e}^{2 \pi {\rm i} \frac{k_1 j_1}{N_x}}$ and
${\rm e}^{2 \pi {\rm i} \frac{k_2 j_2}{N_y}}$ as well as the summation)
gives us the equation (we write $k$ for $k_1 k_2$)
\begin{equation}
\partial_t A_k(t) = (1- B_x k_x^2 - B_y k_y^2) A_k(t) + N_k(t)
\label{CGLeqn:cglkspce}
\end{equation}
where $N_k(t)$ is the Fourier transform of $-C|A|^2 A$ (the nonlinear part of
equation (\ref{CGLeqn:newcgl})).
We simplify this even further by defining
\begin{equation}
L_k \equiv (1-B_x k_x^2 - B_y k_y^2)
\label{CGLeqn:linearpart}
\end{equation}
to give
\begin{equation}
\partial_t A_k(t) = L_k A_k(t) + N_k(t).
\label{CGLeqn:bothparts}
\end{equation}
Multiplying both sides of equation (\ref{CGLeqn:bothparts}) by an
integrating factor ${\rm e}^{-L_k t}$ and simplifying gives
\begin{equation}
\partial_t \left[ {\rm e}^{-L_k t} A_k(t) \right] =
	{\rm e}^{-L_k t} N_k(t).
\label{CGLeqn:collect}
\end{equation}
We then integrate from $t$ to $t+\delta t$ to obtain
\begin{equation}
{\rm e}^{-L_k (t + \delta t)} A_k(t+\delta t) - {\rm e}^{-L_k t} A_k(t) =
	\int_t^{t+\delta t} {\rm e}^{-L_k t'} N_k(t') {\rm d}t'
\label{CGLeqn:dointegral}
\end{equation}
or
\begin{equation}
A_k(t+\delta t) = {\rm e}^{L_k \delta t} A_k(t) +
	{\rm e}^{L_k (t + \delta t)} \int_t^{t + \delta t}
	{\rm e}^{-L_k t'} N_k(t') {\rm d}t'.
\label{CGLeqn:simpintegral}
\end{equation}
Now, consider the integral in equation (\ref{CGLeqn:simpintegral}).
We approximate it as follows:
\begin{equation}
\int_t^{t+\delta t} {\rm e}^{-L_k t'} N_k(t') {\rm d}t' \approx
	N_k(t+\delta t) \int_t^{t+\delta t} {\rm e}^{-L_k t'}{\rm d}t'=
	N_k(t+\delta t)
	\left[\frac{{\rm e}^{-L_k t}-{\rm e}^{-L_k(t + \delta t)}}
	{L_k}\right].
\label{CGLeqn:approxintegral}
\end{equation}
Substituting this into equation (\ref{CGLeqn:simpintegral}) yields
\begin{equation}
A_k(t + \delta t) = {\rm e}^{L_k \delta t} A_k(t) +
	\left[ \frac{{\rm e}^{L_k \delta t} - 1}{L_k} \right]
	N_k(t + \delta t).
\label{CGLeqn:timestep}
\end{equation}
Finally, we use the Adams--Bashforth second--order time step for
$N_k(t+ \delta t)$:
\begin{equation}
N_k(t+ \delta t) = \frac{3}{2} N_k(t) - \frac{1}{2} N_k(t-\delta t).
\label{CGLeqn:adamsbash}
\end{equation}
This gives us the full time step equation
\begin{equation}
A_k(t + \delta t) = {\rm e}^{L_k \delta t} A_k(t) +
	\left[ \frac{{\rm e}^{L_k \delta t} - 1}{L_k} \right]
	\left[\frac{3}{2} N_k(t) - \frac{1}{2} N_k(t-\delta t)
	\right].
\label{CGLeqn:fulltimestep}
\end{equation}
The solution algorithm for equation (\ref{CGLeqn:fulltimestep}) then
consists of the following steps.
\begin{enumerate}
\item Calculate $-C|A|^2 A$ in real ($j_1 j_2$) space.
\item Transform the result of step 1 to Fourier ($k_1 k_2$) space to
	obtain $N_k(t)$.
\item Perform the time step given by equation (\ref{CGLeqn:fulltimestep})
	to obtain $A_k(t + \delta t)$.  Save this result to use
	in calculating  $A_k(t + 2 \delta t)$.
\item Invert $A_k(t + \delta t)$ to obtain $A({\bf x},t+\delta t)$.
\item Save $N_k(t)$ to use in the Adams--Bashforth part of equation
	(\ref{CGLeqn:fulltimestep}) for calculating $A_k(t + 2 \delta t)$.
\end{enumerate}
Analysis of various $k$--space quantities can be performed
between steps 3 and 4 in the algorithm.
For the initial time step we use $N_k(-\delta t) = N_k(0)$.
We also have started our systems with two different types of initial
conditions.  The results do not depend upon which we choose.
One initial condition is to seed the system with two oppositely
charged topological defects.  The other is to begin with random
initial conditions (for instance a uniform distribution
between $-0.1$ and $0.1$ for the field $A$).
After starting, we have to wait for the system
to ``equilibrate'' to a steady--state condition (meaning fluctuations
about an average number of defects).  This takes, for our typical
parameters, on the order of $5,000$ time steps with a step size
of $\delta t = 0.02$.
The steady--state condition is a statistically stationary state,
where averages depend only on differences of space and time
coordinates.

By using a parallel (distributed) FFT, larger systems can be
studied \cite{CGLref:parallel}.  In the parallel algorithm, we
distribute the data for the system over strips ({\it i.e.} the data
for $j_1 = \frac{N_x}{N_P} (n-1) , \frac{N_x}{N_P} n$ and $j_2 = 1,
N_y$ exists on processor $n$, where $N_P$ is the total number of
processors).  The $\frac{N_x}{N_P}$ FFT's for the y ($j_2$) direction
are calculated locally on each processor.  We then perform a global
transpose of the data for $A$ by using collective communication (all
to all) message passing.  Once we have done this, the FFT's for the x
($j_1$) direction are calculated locally on each processor.  Then we
have the result in Fourier space.  To invert the FFT, we just reverse
the process.

\subsection{Improving the Time Step}
\label{CGLsec:timestep}

The result given for equation (\ref{CGLeqn:adamsbash}) represents a
somewhat uncontrolled approximation that was made in the original code
in order to minimize storage requirements.
There is a more accurate
time step (which takes somewhat more storage), which we explain
in this section (see \cite{CGLref:stoer} for a description of
this process).

We wish to find an approximation to
the equation:
\begin{equation}
I \equiv {\rm e}^{L_k(t + \delta t)} \int_t^{t+\delta t} {\rm e}^{-L_k t'}
	N_k(t') {\rm d}t' \equiv
	{\rm e}^{L_k(t + \delta t)} \int_{t_{p-j}}^{t_{p+l}} {\rm e}^{-L_k t'}
	N_k(t') {\rm d}t'
\label{CGLeqn:abintegral}
\end{equation}
where $t_p = p \delta t = t$, $t_{p-j} = t_p -j \delta t$, and
$t_{p+l} = t_p + l \delta t$.
We then approximate $N_k(t')$ using an interpolating polynomial
\begin{equation}
N_k(t') \approx \sum_{i=0}^q N_k(t_{p-i}) L_i(t'),
\label{CGLeqn:nkapprox}
\end{equation}
where
\begin{equation}
L_i(t') = \prod_{l=0,l \neq i}^q \frac{t-t_{p-l}}{t_{p-i}-t_{p-l}}
\label{CGLeqn:interp}
\end{equation}
is the interpolating polynomial.
Substituting this into (\ref{CGLeqn:abintegral}) gives
\begin{equation}
I = {\rm e}^{L_k(t+\delta t)}
	\sum_{i=0}^q N_k(t_{p-i}) \int_{t_{p-j}}^{t_{p+l}}
	{\rm e}^{-L_k t'} L_i(t') {\rm d}t'.
\label{CGLeqn:abintsubs}
\end{equation}
For comparison with the standard Adams--Bashforth formula, we are interested
in the case $j=0$, $l=1$, and $q=1$.   Other choices for $j$, $l$, and $q$
give other time--stepping schemes ({\it e.g.} Crank--Nicholson is
given by $j=1$, $l=0$, and $q=1$).
For our case the interpolating polynomials are
\begin{equation}
L_0(t') = \frac{t'-t_{p-1}}{t_p - t_{p-1}} =
	\frac{t'-t + \delta t}{\delta t}
\label{CGLeqn:lzero}
\end{equation}
and
\begin{equation}
L_1(t') = \frac{t'-t_{p}}{t_{p-1} - t_p} =
	\frac{t'-t}{- \delta t}.
\label{CGLeqn:lone}
\end{equation}
Using these results we obtain:
\begin{equation}
\int_{t_p}^{t_{p+1}} {\rm e}^{-L_k t'} L_0(t') {\rm d}t' =
	\frac{{\rm e}^{-L_k t}}{L_k} \left[ {\rm e}^{-L_k \delta t}
	\left(-2 -\frac{1}{L_k \delta t} \right) +
	\left( 1 + \frac{1}{L_k \delta t} \right) \right]
\label{CGLeqn:lzeroint}
\end{equation}
and
\begin{equation}
\int_{t_p}^{t_{p+1}} {\rm e}^{-L_k t'} L_1(t') {\rm d}t' =
	- \frac{{\rm e}^{-L_k t}}{L_k} \left[ {\rm e}^{-L_k \delta t}
	\left(-1 -\frac{1}{L_k \delta t} \right) +
	\frac{1}{L_k \delta t} \right].
\label{CGLeqn:loneint}
\end{equation}
The full integral (\ref{CGLeqn:abintegral}) becomes
\begin{eqnarray}
I & = & N_k(t) \left[ \frac{1}{L_k} \left( -2 -\frac{1}{L_k \delta t} \right)
	+ \frac{{\rm e}^{L_k \delta t}}{L_k} \left( 1 + \frac{1}{L_k \delta t}
	\right) \right] \nonumber \\[10pt]
 & & - N_k(t -\delta t) \left[ \frac{1}{L_k} \left(-1-\frac{1}{L_k \delta t}
	\right) + \frac{{\rm e}^{L_k \delta t}}{L_k^2 \delta t} \right].
\label{CGLeqn:abintlast}
\end{eqnarray}

In the limit $L_k \rightarrow 0$, equation (\ref{CGLeqn:lzeroint})
goes to $\frac{3}{2} \delta t$ and equation (\ref{CGLeqn:loneint})
goes to $-\frac{1}{2} \delta t$, which are just the results for the
standard Adams--Bashforth second--order formula.  Also,
when $L_k$ is held fixed and $\delta t \rightarrow 0$, the leading
order terms become $\frac{3}{2} \delta t$ and $\frac{1}{2} \delta t$,
with corrections of $O(\delta t^2)$, so that the original time step
(equation (\ref{CGLeqn:fulltimestep}))
is still ``correct'' to $O(\delta t)$.  However, the improved
time step developed in this section allows simulations
to be run with a larger time increment $\delta t$ for the same
desired numerical accuracy.  This time step method has been
used for simulations of Rayleigh--B\'enard convection
\cite{CGLref:Pesch}.  Our results were obtained with the
unimproved time step, but we present the improved version here
so that others may take advantage of it.

\section{Results for $A({\bf x},t)$}
\label{CGLsec:Aresults}

We present some results for the field $A({\bf x},t)$.  First,
Figure \ref{CGLfig:AAcorr} shows that the field--field
correlation function decays exponentially (or perhaps faster),
and shows no hints of long--range order (such as a power law decay might
suggest).
We also note that the x and y directions show statistically
different behavior.  The y direction appears to show a domain type
structure (as in, for example, spinodal decomposition \cite{CGLref:Spinodal})
for $r \approx 10$, because of the significant amount of
negative correlations in that region.  The x direction does not
appear to show this behavior, as it decays smoothly to zero
from its value at $r=0$.

The exponential decay of the $A$--$A$ correlation function
has been observed in previous simulations
\cite{CGLref:defturbI,CGLref:defturbII}, and it was suggested by
these authors that
the defects are responsible for the strong decorrelation of the
field.  We can merely note here that the field $A({\bf x},t)$ is
not conserved in time.  This lack of a conservation law
would imply, within the framework of generic scale invariance
discussed below in Section \ref{CGLsec:gsi}, that the field--field
correlation function should decay exponentially.

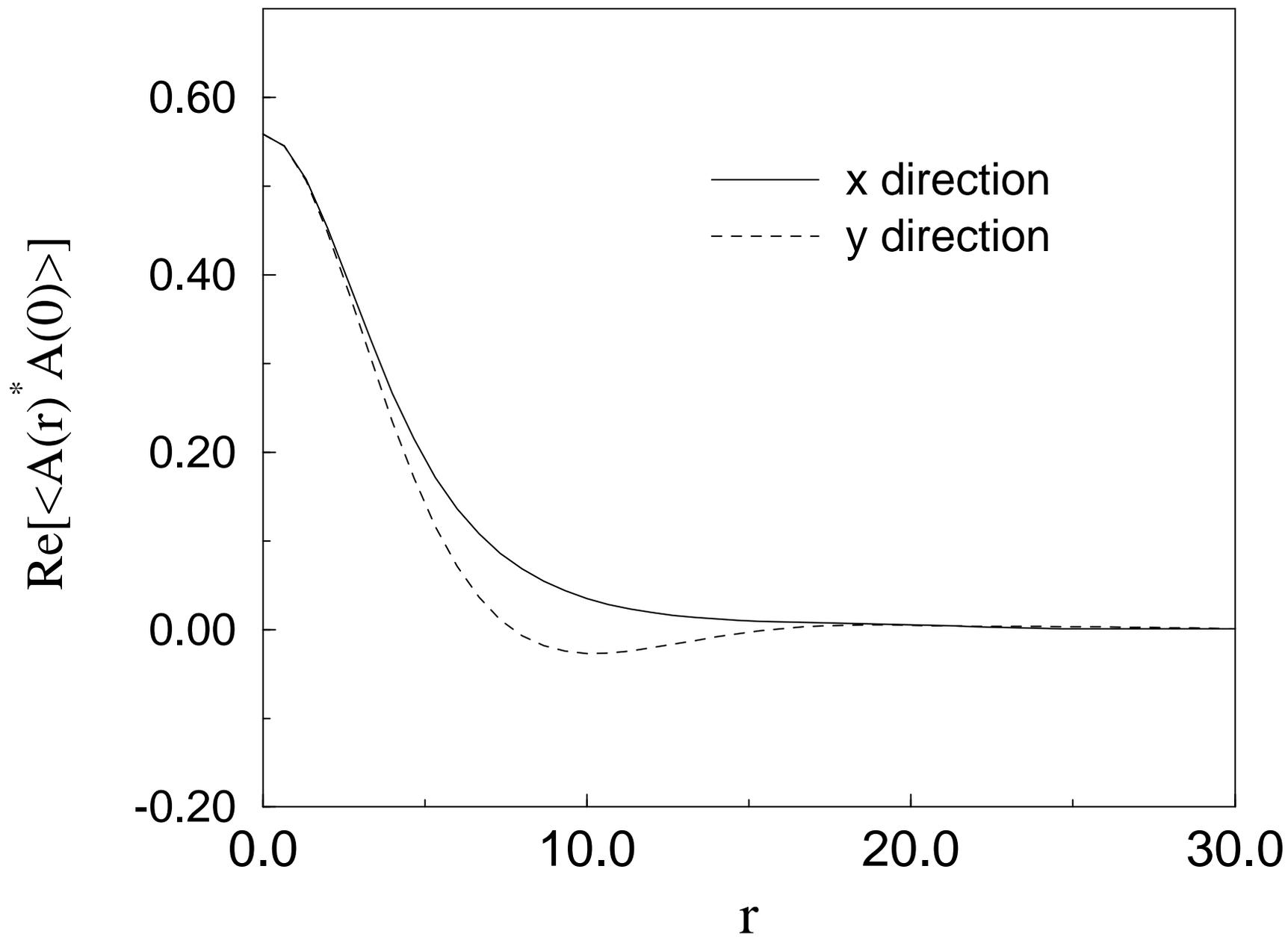
\begin{figure}
\caption[{$\langle S(k_x) \rangle = {\rm Re} \langle A^*(k_x) A(0)\rangle $ }]
{ $\langle S(k_x)\rangle = {\rm Re} \langle A^*(k_x) A(0)\rangle $
averaged over $500,000$ time steps, with parameter
values $c = 1.5$, $b_x = -1.5$, and $b_y = -1.5$.  The
system has size $L_x = L_y = 60$ with $90$ Fourier modes in each direction.
Other Fourier modes show similar behavior.
}
\label{CGLfig:avgSk}
\end{figure}

In Figure \ref{CGLfig:avgSk} we show the averaged power
spectrum $\langle S(k_x)\rangle={\rm Re}\langle A^*(k_x)A(0)\rangle$;
results are similar for $\langle S(k_y) \rangle$.
This falloff is exponential (or perhaps faster), which is to be
expected because of the fast decay of the field--field correlation
function presented in Figure \ref{CGLfig:AAcorr}.
The fact that the power spectrum decays rapidly to zero means that
we don't have problems with aliasing of the Fourier transform in
the numerical method.

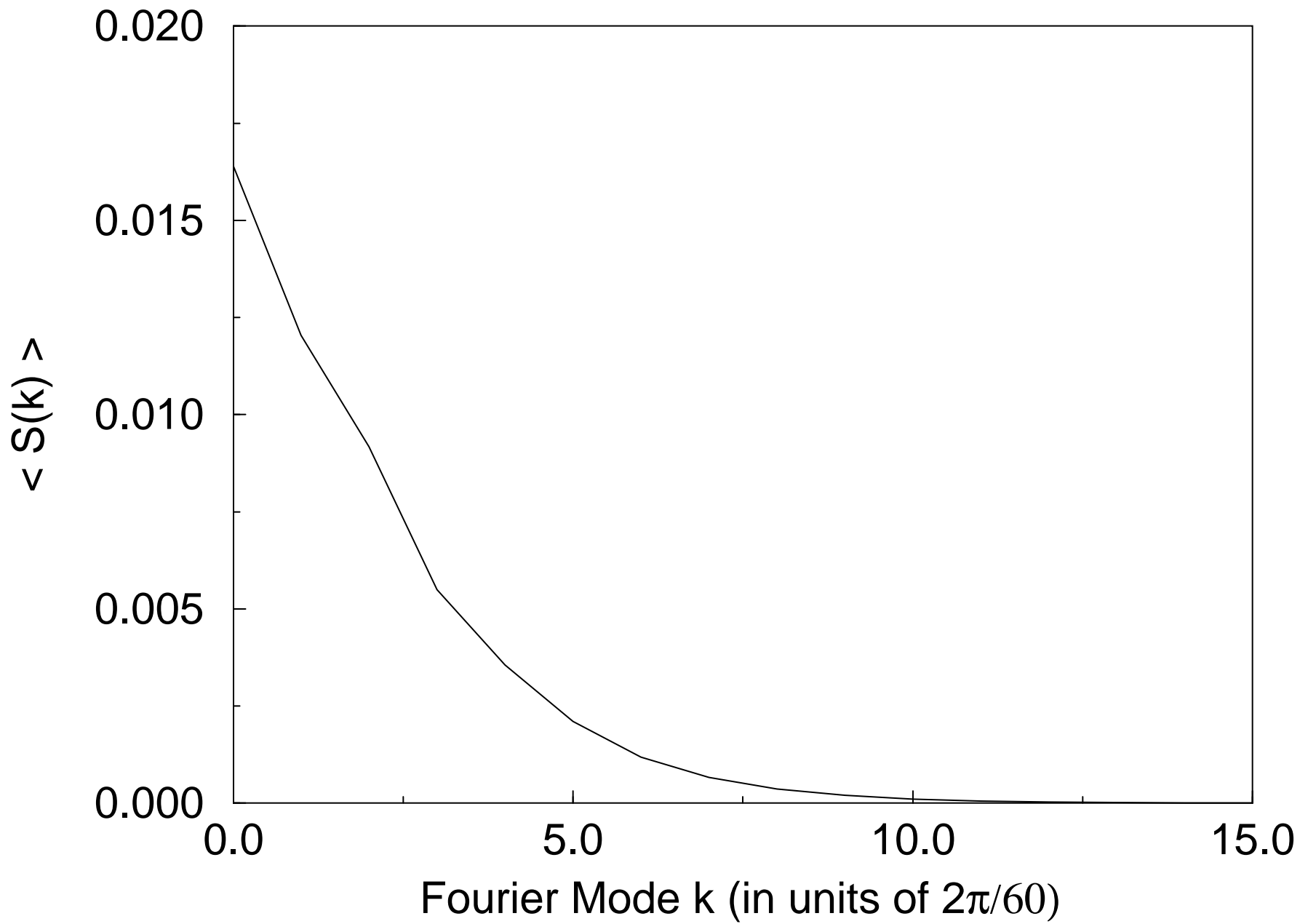
\begin{figure}
\caption{Prob(${\rm Re}~[A(x)]$) averaged over space at
a particular time (solid line) and over time at a particular spatial
location (dashed line).   The spatial average is over 129,600 points,
while the time average is over 200,000 time steps.
This is for $c = 1.5$, $b_x = -0.75$, and $b_y = -3.0$.
}
\label{CGLfig:probAx}
\end{figure}

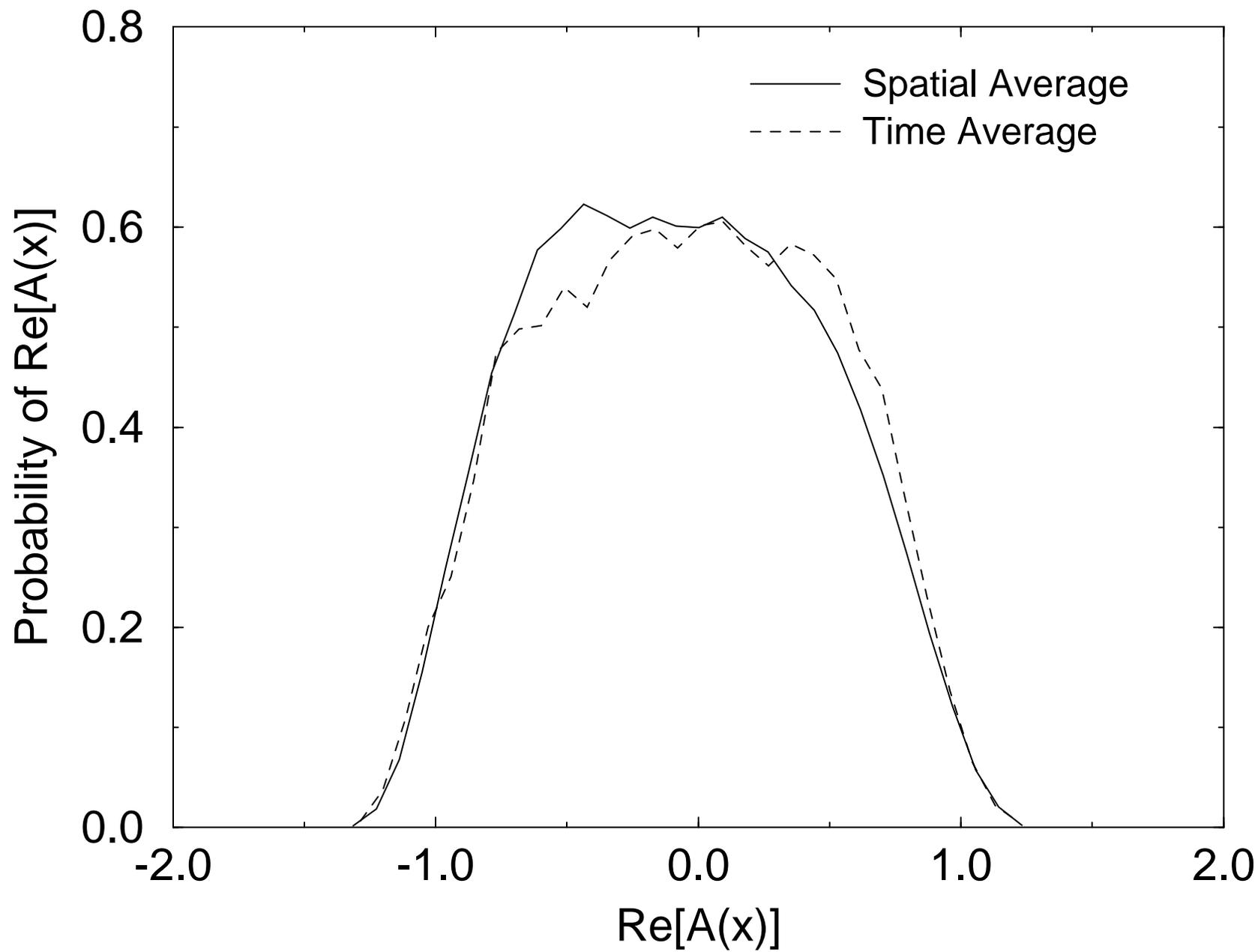
\begin{figure}
\caption{Prob($|A(x)|$) averaged over space at
a particular time (solid line) and over time at a particular spatial
location (dashed line).   The spatial average is over 129,600 points,
while the time average is over 200,000 time steps.  Note that
values of $|A|>1.0$ represent shocks in the system, while
the nonzero values at the limit $|A|=0$ represent the defects in the
system. This is also for $c = 1.5$, $b_x = -0.75$, and $b_y = -3.0$.
}
\label{CGLfig:probabsAx}
\end{figure}

The probability of a particular
value of ${\rm Re} [A(x)]$ versus ${\rm Re} [A(x)]$ is shown in
Figure \ref{CGLfig:probAx}.  This figure includes both
spatial averaging at a particular time, as well as
time averaging at a particular point in order to test
the system's ergodicity.
The plot shows no special structure,
and is not expected to, because principles that might give
a special form, such as the Central Limit
Theorem, do not apply to the measured quantity.
Similar results are obtained for the probability of
${\rm Im} [A(x)]$.  We note that the two methods of averaging
agree fairly well.  We would expect that for larger systems and longer times
the two results should agree.
In Figure \ref{CGLfig:probabsAx} we present the
probability of a particular value of $|A(x)|$ versus
$|A(x)|$, again with both types of averaging.
Values of $|A|$ greater than $1.0$ represent
shock waves in the system, and the probability of larger and larger
values of $|A|$ should decrease above $|A|=1.0$
When $|A|=0$, there is a defect in the system.  This is expected to
be rarer than other nonzero values, because it requires
both ${\rm Re} A = 0$ and ${\rm Im} A = 0$.  Both of these
expected results are seen in the data.

In Figure \ref{CGLfig:probAk} we show the result for the probability
of $A(k)$ versus $A(k)$.  Non--Gaussian fluctuations in Fourier
variables have been measured in capillary wave fields, and this has been
used to question the applicability of a thermodynamic description
for spatiotemporal chaos \cite{CGLref:faradayint}.
In contrast to the results for $A(x)$, we expect the Central Limit
Theorem to hold for $A(k)$.  This would then predict
that the probability distribution function would be a Gaussian.
In Figure \ref{CGLfig:probAk} we compare
our results to a Gaussian, and we see that we certainly have Gaussian
fluctuations.  Similar results hold for other wavevectors $k$ for which
we have measured this quantity.
We also measured this probability distribution for these fluctuations
for a different set of parameters ($c=-0.5$, $b_x = b_y = 40.0$)
and started with a small system.  The small system showed non--Gaussian
fluctuations.  When we went to larger systems, the fluctuations became
Gaussian.  This just means that for the small system there were not
enough underlying degrees of freedom for
the central limit theorem to be valid when we averaged in the system.
Note that this small system also showed quasiperiodic behavior in
the time series of $A({\bf x},t)$ for fixed ${\bf x}$.
This system had $L_x = L_y = 60$.  When we increased the size of the
system to $L_x = L_y = 120$, this quasiperiodic behavior disappeared and
was replaced by the chaotic behavior seen in \cite{CGLref:aransoninteractI}.
This larger system was the one that also showed the Gaussian fluctuations.

Time averages of spatiotemporal chaos in experimental systems result
in periodic spatial structures \cite{CGLref:averaging,CGLref:averaging2}.
We also tried averaging our amplitude $A({\bf x},t)$ over various
times, and we did not see any particular persistent structures
emerging.  These results were not, however, from extensive tests.  It
is likely that the experimental results are due to the fixed boundary
in these systems, which constrains the system to fluctuate about
certain well--defined normal modes.

All of the above results for the field itself were not
encouraging from the standpoint of illuminating special features of
spatiotemporal chaos.
We wish to examine another possible approach
for describing the system, namely, generic scale invariance.

\section{Generic Scale Invariance}
\label{CGLsec:gsi}

In equilibrium systems, spatial correlations typically decay
exponentially.  For non\-equilibrium systems, such as those with an
external driving force, the situation can be quite different.  For
a nonequilibrium system with a conservation law and external noise,
spatial correlation functions can decay algebraically.  It has been
suggested that this algebraic decay is expected to occur for a broad
range of conditions, and this has been called ``generic scale
invariance''
\cite{CGLref:gsiI,CGLref:gsiII,CGLref:gsiIII,CGLref:grinsand,CGLref:garrido}.
Some extended deterministic chaotic
systems also exhibit algebraic
decay \cite{CGLref:cmapl,CGLref:solvedchaos,CGLref:cellular}.  In at
least one of these examples, the chaotic fluctuations appear to play
the same role as stochastic noise \cite{CGLref:cmapl}.  The complex
Ginzburg--Landau equation would seem to satisfy the criterion for
generic scale invariance;  it shows nonequilibrium behavior since it
cannot be derived from an underlying potential ({\it i.e.} it is
non--relaxational), and in a system with periodic boundary conditions,
the topological order parameter $\rho({\rm \bf r})$ is conserved.
Finally, we conjecture that the chaotic noise in our system plays the
role of stochastic noise, again for the purpose of testing the
applicability of generic scale invariance.

With this set of conditions, we could have a hydrodynamic equation
for the conserved order parameter (coarse grained over distances larger
than the typical spacing between defects):
\begin{equation}
\partial_t \rho({\rm \bf r},t) = \Gamma\{ \rho({\rm \bf r},t) \} +
	\eta({\rm \bf r},t)
\label{CGLeqn:gsi}
\end{equation}
where $\Gamma$ is a general conserving operator on $\rho$,
such as $\Gamma_0 \nabla^2 + \Gamma_1 (\nabla^2)^2 + \Gamma_{2x}
\partial_x^4 + \Gamma_{2y} \partial_y^4$.  It can also
contain nonlinear terms ({\it e.~g.}
$\nabla \cdot [(\nabla^2 \rho) (\nabla \rho)]$ ) .
The stochastic noise term $\eta$ is determined by:
\begin{equation}
\langle \eta({\rm \bf r},t) \rangle = 0
\label{CGLeqn:noisemean}
\end{equation}
\begin{equation}
\langle \eta({\rm \bf r},t) \ \eta({\rm \bf r}~',t') \rangle =
	D \delta({\rm \bf r} - {\rm \bf r}~')\delta(t-t'),
\label{CGLeqn:noisestddev}
\end{equation}
where $D$ must be composed of
differential operators for our strictly
conserved order parameter.
This conserving noise term represents the effect of the
chaotic fluctuations in the complex Ginzburg--Landau equation.
There is evidence from the mapping of the Kuramoto--Sivashinsky equation
to the Kardar--Parisi--Zhang equation
\cite{CGLref:kstokpz,CGLref:kstokpzcomment,CGLref:kstokpzreply}
and from coupled
map lattices \cite{CGLref:cmapl} that this identification of spatiotemporal
chaotic fluctuations with stochastic noise is not unreasonable.
Note also that we are assuming that we can write down a local equation of
motion for the order parameter $\rho$.

For systems with nonconserving noise ({\it i.e.} $D$ is a constant),
equation (\ref{CGLeqn:gsi}) is expected to always give rise to power
law decays in the two point correlation function $G_\rho({\rm \bf r})
\equiv \langle \rho({\rm \bf r}) \rho({\rm \bf 0}) \rangle$,
as well as in higher order correlation functions.  For nonlinear systems with
conserving noise ({\it e.~g.} $D = D_1 \nabla^2$) the situation is
somewhat more complicated \cite{CGLref:gsiII}.
If the system is isotropic, then one obtains exponential decays in
$G_\rho({\rm \bf r})$, but power law decays occur in higher order
correlation functions.  Systems which break isotropy give rise to
algebraic decay in $G_\rho({\rm \bf r})$.  For systems with cubic
symmetry, one expects $G_\rho({\rm \bf r}) \sim 1/r^{d+2}$ for large
$r$.  For systems which break cubic symmetry one expects $G_\rho({\rm
\bf r}) \sim 1/r^d$.  We will study the last regime, a
two--dimensional system with broken square symmetry, where generic
scale invariance predicts that
\begin{equation}
G_\rho^{generic}({\rm \bf r}) =
	\langle \rho({\rm \bf r}) \rho({\rm \bf 0}) \rangle \sim 1/r^2
\label{CGLeqn:gingsi}
\end{equation}
for large $r$ \cite{CGLref:gsiII}.
We shall compare this prediction to the results from our numerics.

The ideas behind generic scale invariance depend upon
showing that nonlinearities in the equations
are irrelevant in a renormalization group sense.
This means that as the scale on which the system is examined
grows, the importance of the nonlinearities in the dynamics at the
larger scale is less important.  This analysis is usually carried out
in a perturbative manner, where the strength of the nonlinear
term is the small expansion parameter.
It has proven notoriously difficult to treat
topological defects in a perturbative manner.
An example of this is the Kosterlitz--Thouless transition \cite{CGLref:kt}.

We also note that generic scale invariance requires short--ranged
interactions.  There is some evidence that this occurs
for the defects in the complex Ginzburg--Landau equation
\cite{CGLref:Interactions,CGLref:araninter},
but collective effects might be important.
Finally, the mapping of chaotic fluctuations to the
stochastic noise could break down. In the next section, we will
prove that the predictions of generic scale invariance
cannot apply to topological defects, and therefore at least one of the
requirements for it to be present is lacking in our system.

\section{Topological Constraints}
\label{CGLsec:constraints}

We define the excess order parameter in a region to be
\begin{equation}
\delta \rho_L \equiv \left| \int_{{\rm \bf r} \in B(L)}
	\left( \rho({\rm \bf r}+ {\rm \bf r}_0) -
	\rho_0 \right) {\rm d \bf r}  \right|,
\label{CGLeqn:deltarho}
\end{equation}
where $B(L)$ represents a circle of radius $L$ about a point ${\rm \bf r}_0$,
which we take as ${\rm \bf r}_0 = {\rm \bf 0}$ due to translational
invariance, and where $\rho_0$ is the average order parameter (in our case 0).
For nontopological objects, the constraint is given
by $\delta \rho_L \leq a_1 L^2$, where $a_1$ is some numerical constant.
The excess of a nontopological object in a particular region must scale
as the area of that region, since each individual object occupies
a fixed area.
For topological objects of the type we are studying,
this constraint is different, {\it i.~e.}
$\delta \rho_L \leq a_2 L$, where $a_2$ is again some numerical
constant.  The constraint arises from the fact that any excess
of topological defects in a region must be detectable  simply by
traversing the perimeter of that
region.  Each topological defect has
an ``arm'' with characteristic width that
must pass through the perimeter of the region.  Examples of this
are the spiral arms of the defects in Rayleigh--B\'enard convection,
extra rows of atoms for dislocations in crystals, and in our case
lines of ${\rm Re}~[A] = 0$ and ${\rm Im}~[A] = 0$.
When a region contains the maximum excess number of defects allowed,
each of these lines takes up a fixed amount of the perimeter of the region.
Since the
maximum excess number of topological objects scales linearly with the
number of lines, and the number of lines scales as the perimeter of
the region, we must have that the maximum excess number of topological
objects scales as the linear size $L$ of the region.

If we assume that the correlation function $G_\rho({\rm \bf r})$
decays at the same asymptotic rate independent of the direction of
${\rm \bf r}$, {\it i.e.}  $G_\rho({\rm \bf r}) \sim f(\theta) g(r)$
where $g(r) \sim 1/r^\alpha$ for large $r$, then with this constraint
we can show for two dimensions that $\alpha$ must be greater than 2,
which contradicts the prediction of generic scale invariance.
This result also requires that $\int_0^{2 \pi} f(\theta) d\theta \neq
0$, which we expect to be true except for special cases;
we discuss this point further in Section \ref{CGLsec:splitting}
below.
A correlation function satisfying both assumptions occurs, for example,
in studies of non--equilibrium conservative anisotropic lattice gases
\cite{CGLref:kawasaki,CGLref:lebowitz}.
To show that $\alpha > 2$,
we use the inequality $\delta \rho_L \leq a L$.
Squaring this relation yields
\begin{equation}
\int_{{\bf r} \in B(L)} \int_{{\bf r}' \in B(L)} {\rm d\bf r} {\rm d\bf r}'
	\rho({\bf r},t) \rho({\bf r}',t) \leq a^2 L^2.
\label{CGLeqn:appsqrdelta}
\end{equation}
Now we average over the noise (or over space and time, depending upon
how you wish to look at it)
\begin{equation}
\langle \rho({\bf r},t) \rho({\bf r}',t) \rangle
	\rightarrow G_\rho({\bf r} - {\bf r}').
\label{CGLeqn:appavgnoise}
\end{equation}
This then gives the constraint equation
\begin{equation}
\langle \delta \rho_L^2 \rangle ~ \leq ~ a^2 L^2
\label{CGLeqn:avgconstraint}
\end{equation}
We now make the assumption that we can write $G_\rho({\bf r})$ in the
form $f(\theta) g(r)$ where $\theta$ is the angle for ${\bf r}$ and
$r = |{\bf r}|$.  Strictly speaking, this assumption is only necessary
for large $r$.
Given that $G_\rho$ takes this form, we
can write
\begin{equation}
\int_{{\bf r} \in B(L)} \int_{{\bf r}' \in B(L)} {\rm d \bf r}
	{\rm d \bf r}'
	G_\rho({\bf r} - {\bf r}') =
	\frac{1}{2 \pi} \int_0^{2\pi} f(\theta) {\rm d}\theta
	\int_0^{2L} g(R) w(R) {\rm d}R,
\label{CGLeqn:appweight}
\end{equation}
where $w(R)$ is given by
\begin{equation}
w(R) = \int_{{\bf r} \in B(L)} \int_{{\bf r}' \in B(L)}
	S(R) \delta({\bf r} - {\bf r}' - {\bf R}) {\rm d \bf r}
	{\rm d \bf r}'
\label{CGLeqn:appweightdef}
\end{equation}
with $S(R) = 2 \pi R$ the surface area of the circle of radius $R$.
We further will assume that $\int_0^{2\pi} f(\theta) {\rm d}\theta \neq 0$,
and now calculate a closed form expression for $w(R)$.
To do this we use the definition of $\delta({\bf x})$:
\begin{equation}
\delta({\bf x}) \equiv \frac{1}{(2\pi)^2} \int_{-\infty}^{\infty}
	{\rm d \bf k} \ {\rm e}^{{\rm i} {\bf k} \cdot {\bf  x}}.
\label{CGLeqn:appdeltadef}
\end{equation}
Plugging this into equation (\ref{CGLeqn:appweightdef}) gives
\begin{equation}
w(R) = \frac{2\pi R}{(2\pi )^2} \int_{-\infty}^{\infty} {\rm d \bf k} \
	{\rm e}^{-{\rm i} {\bf k} \cdot {\bf R}} \int_{{\bf r} \in B(L)}
	{\rm d \bf r} \
	{\rm e}^{{\rm i} {\bf k} \cdot {\bf r}} \int_{{\bf r}' \in B(L)}
	{\rm d \bf r}' \ e^{-{\rm i} {\bf k} \cdot {\bf r}'}.
\label{CGLeqn:appwrsub}
\end{equation}
The integrals over $B(L)$ are given by
\begin{equation}
\int_{{\bf r} \in B(L)} {\rm d \bf r} \ {\rm e}^{{\rm i}
	{\bf k}\cdot {\bf r}} =
	\int_0^L \int_0^{2\pi} r {\rm d}r {\rm d}\theta
	{\rm e}^{{\rm i} k r \cos \theta} =
	\frac{2 \pi L}{k} J_1(k L),
\label{CGLeqn:appintb}
\end{equation}
where $J_1$ is the first order Bessel function.
Substituting this result in equation (\ref{CGLeqn:appwrsub}), we obtain
\begin{equation}
w(R) = 2 \pi R L^2 \int_{-\infty}^{\infty} {\rm d \bf k}
	{\rm e}^{-{\rm i} {\bf k} \cdot {\bf R}}
	\frac{J_1^2(k L)}{k^2}
     = 4 \pi^2 R L^2 \int_0^\infty \frac{{\rm d} k}{k}
	J_0\left(k \frac{R}{L}\right) J_1^2(k).
\label{CGLeqn:appwrbessel}
\end{equation}
Performing this last integral \cite{CGLref:prudnikov} gives
\begin{equation}
w(R) = 4 \pi R L^2 \left[ \cos^{-1}\left(\frac{R}{2L}\right) -
	\frac{R}{2L} \sqrt{1-\left(\frac{R}{2L}\right)^2} \right].
\label{CGLeqn:appwrfinal}
\end{equation}
We can check this by noting that $\int_0^{2 L} w(R) {\rm d}R = \pi^2 L^4$,
which is just $\left(\int_{{\bf r} \in B(L)} {\rm d \bf r}\right)^2$.

Next, we apply $w(R)$ to the problem at hand.  Suppose that
$g(R)$ in equation (\ref{CGLeqn:appweight}) exhibits its asymptotic behavior
outside of some $R = r_{min}$, $g(R) \sim R^{-\alpha}$ for $R > r_{min}$
and also that $L \gg r_{min}$.  Splitting the
integral into two parts yields
\begin{equation}
\int_0^{2 L} g(R) w(R) {\rm d}R = \int_{r_{min}}^{2 L} + \int_0^{r_{min}}
	g(R) w(R) {\rm d}R.
\label{CGLeqn:appsplit}
\end{equation}
We consider each of these terms separately.  The second term
can be bounded in the following manner:
\begin{equation}
\left| \int_0^{r_{min}} g(R) w(R) {\rm d}R \right| \leq |g_{max}|
	\int_0^{r_{min}} w(R) {\rm d}R.
\label{CGLeqn:appgmax}
\end{equation}
Expanding the integral of $w(R)$ in descending powers of $L$ gives
\begin{equation}
\left| \int_0^{r_{min}} g(R) w(R) {\rm d}R \right| \leq \pi^2 \ r_{min}^2 \
	g_{max} \ L^2 + {\rm O}(L).
\label{CGLeqn:appgmaxfinal}
\end{equation}

We calculate the first integral in equation (\ref{CGLeqn:appsplit}) next.
Assuming that $g(R) = g_0 R^{-\alpha}$, we substitute equation
(\ref{CGLeqn:appwrfinal}) for $w(R)$ and write the integral as
\begin{equation}
4 \pi L^2 (2 L)^{2-\alpha} g_0 \int_{\frac{r_{min}}{2L}}^1
	\left[\eta^{1-\alpha} \cos^{-1}\eta - \eta^{2-\alpha}
	(1-\eta^2)^{\frac{1}{2}} \right] {\rm d}\eta.
\label{CGLeqn:appsubs}
\end{equation}
We again expand this integral in descending powers of $L$.
There are no singularities due to the upper limit of $1$, since the
integrand is analytic there for all $\alpha$.
However, at the lower limit, as $L \rightarrow \infty$, there can be
divergences, depending on the value of $\alpha$.
The integrand, upon expansion about $\eta = 0$, becomes
\begin{equation}
\eta^{-\alpha} \left[ \frac{\pi}{2} \eta - 2 \eta^2
	+ {\rm O}(\eta^4) \right].
\label{CGLeqn:appexpand}
\end{equation}
Now, suppose that $\alpha < 2$.  Then the integrand has an integrable
singularity at $\eta = 0$, and as $L \rightarrow \infty$ we obtain
some constant plus correction terms that die away for large $L$.
For $\alpha = 2$ we get a logarithmic divergence, and for
$\alpha > 2$ a diverging term $L^{\alpha-2}$ which exactly
cancels the $L^{2-\alpha}$ in front of equation (\ref{CGLeqn:appsubs}).

The leading order
behavior of $\langle \delta \rho_L^2\rangle$ is then
\begin{eqnarray}
\alpha < 2 & : & \langle \delta \rho_L^2 \rangle
	\sim L^2 L^{2-\alpha} \nonumber \\
\alpha = 2 & : & \langle \delta \rho_L^2 \rangle
	\sim L^2 \log(L) \nonumber \\
\alpha > 2 & : & \langle \delta \rho_L^2 \rangle
	\sim L^2.
\label{CGLeqn:alpha}
\end{eqnarray}
Recall that the result for generic scale invariance is the case
$\alpha = 2$ (see equation (\ref{CGLeqn:gingsi})).  However, this case
violates the constraint given by equation (\ref{CGLeqn:avgconstraint}).
Within our assumptions, this result means that for topological objects
the results of generic scale invariance cannot hold.  In fact, what we
have provided is a bound on $\alpha$.  For topological objects,
$\alpha$ must be strictly greater than $2$.  Generic scale invariance
predicts $\alpha = 2$.  The simple geometric nature of topological
objects prevents them from having correlation functions that decay as
certain power laws.  (A decay of $1/(r^2 (\log r)^\beta)$ for $\beta > 1$
satisfies our bounds on $\langle \delta \rho_L^2 \rangle$, but
contradicts the predictions of generic scale invariance.)
An added conclusion from the consideration of the topological
constraints is that if the topological
objects form ordered states, they must be of the antiferromagnetic
variety ({\it e.g.} alternating $+$ and $-$ vortices) in at least one
direction, in order to satisfy the topological constraint.  An example
of such a state in the complex Ginzburg--Landau equation has been
seen, with defects ordering along chains where the defects in one
chain are of opposite sign from the defects in neighboring chains
\cite{CGLref:advmat}.

In our analysis we have only considered the largest possible
fluctuations.  We expect these fluctuations to be rare, and hence
expect a faster decay than the bound we provide.  As an analogy, for
nontopological objects the analysis presented here would predict that
the correlation function can be at most a constant for large r; in
practice, systems like spins or atoms have connected correlation
functions that decay to zero, either as power laws or as exponentials.
However, it is often true that inequalities in physics are saturated,
especially in relationships between critical exponents for phase
transitions.  This could be the case here, but numerically it appears
that we do not saturate this bound, as will be shown in Section
\ref{CGLsec:numerics} below.

\subsection{Splitting the Correlation Function}
\label{CGLsec:splitting}

One further point that we should consider is the splitting of the
correlation function into angular and radial components.
We mentioned above that this is important for our analysis.
We cannot explicitly show this for the topological defects, but we
can demonstrate this splitting in the context of generic scale invariance.
Starting with the general result for the correlation function
\begin{equation}
G({\bf r}) = \int {\rm d \bf k} {\rm e}^{{\rm i} {\bf k} \cdot {\bf x}}
	\frac{D({\bf k})}{\Gamma({\bf k})}
\label{CGLeqn:gsicorr}
\end{equation}
we then define
\begin{equation}
G(r) = \frac{1}{2 \pi} \int_0^{2 \pi} G({\bf r}) {\rm d} \theta
\label{CGLeqn:thetacorr}
\end{equation}
where ${\bf r} = (r \cos \theta, r \sin \theta)$.
Then
\begin{equation}
G(r) = \frac{1}{2 \pi} \int_0^{2 \pi} {\rm d} \theta \int {\rm d}
	 {\bf k} \frac{D({\bf k})}{\Gamma({\bf k})}
	{\rm e}^{{\rm i}(k_x r \cos \theta + k_y r \sin \theta)}.
\label{CGLeqn:subscorr}
\end{equation}
Performing the $\theta$ integral yields
\begin{equation}
G(r) = \int {\rm d \bf k} \frac{D({\bf k})}{\Gamma({\bf k})}
J_0 ( r \sqrt{ k_x^2 + k_y^2 }),
\label{CGLeqn:corrbessel}
\end{equation}
where $J_0$ is the zeroth order Bessel function.  Note that this
expression is not in general zero.

We can give a concrete expression, in
a particular limit, for $G(r)$.  We begin with specific expressions
for $D({\bf k})$ and $\Gamma({\bf k})$, and ignore
questions of convergence of the integrals for large $k$.
Consider
\begin{equation}
G({\bf r}) = \int_{-\infty}^{\infty} \int_{-\infty}^{\infty}
	{\rm d}k_x {\rm d}k_y {\rm e}^{{\rm i}(k_x x + k_y y)}
	\left( \frac{a k_x^2 + b k_y^2}{c k_x^2 + d k_y^2 } \right).
\label{CGLeqn:plugin}
\end{equation}
The $k_x$ integral can be performed as a contour integral; the
expression has poles at
$k_x = \pm i \sqrt{d/c} k_y \equiv \pm i \gamma k_y$.
We assume for ease of calculation the $x > 0$ and $y > 0$; the
results are the same for other cases.
When $k_y > 0$, we close the
contour in the upper half of the complex $k_x$ plane,
enclosing the pole at $i \gamma k_y$.
For $k_y < 0$ we enclose the pole at $ - i \gamma k_y$.
The integral then becomes $2 \pi i$ times the residue at
$k_x = i \gamma k_y$. Then we obtain
\begin{equation}
G({\bf r}) = \frac{\pi}{c \gamma} (b - a \gamma^2) \left(
	\int_0^{\infty} {\rm d}k_y~k_y {\rm e}^{-k_y (\gamma x - {\rm i} y) }
	+ \int_{-\infty}^{0} {\rm d}k_y~k_y {\rm e}^{-k_y (-\gamma x -
	{\rm i} y)} \right).
\label{CGLeqn:firstintegral}
\end{equation}
By performing the $k_y$ integrals we obtain (using $x = r \cos \theta$ and
$y = r \sin \theta$)
\begin{equation}
G({\bf r}) = \frac{2 \pi (b c - a d)}{\sqrt{ c d }}
	\left( \frac{ d~\cos^2\theta - c~\sin^2\theta }{
	(d~\cos^2\theta + c~\sin^2\theta)^2} \right) \frac{1}{r^2}
\label{CGLeqn:finalcorr}
\end{equation}
Note that our results are for the case $c \neq d$, so we have for
this simple example the result that the angular average is
not zero.  Also, equation (\ref{CGLeqn:finalcorr}) is explicitly
split into a radial and angular component.

\section{Numerical Results for the Defects}
\label{CGLsec:numerics}

We have numerically solved equation (\ref{CGLeqn:cgl}) with periodic
boundary conditions in the turbulent regime using a
Fourier collocation code.
We choose $L_x = 240$, $L_y = 240$, $N_x = 360$, and
$N_y = 360$ (see Section \ref{CGLsec:nummethod} above).
We use the parameter values $c = 1.5$, $b_x = -0.75$, and $b_y = -3.0$.
The time step used was
$\delta t = 0.02$.  We initially ``equilibrate'' from a state with two
oppositely charged defects to a state with fluctuations about some
average number of defects.  This typically takes $5000$ time steps.
We note that the defects do not form bound pairs.  When a pair is
created, the defects tend to move apart, and when they eventually
annihilate, they usually do so with a defect other than their initial
partner.  This illustrates that the system is not in a Kosterlitz--Thouless
bound pair phase \cite{CGLref:kt}.  Figure \ref{CGLfig:snapshot}
shows a snapshot of part of our system.

Topological defects have been studied in systems undergoing phase ordering.
In these systems (an example is the XY model) the existence of an underlying
Hamiltonian allows analytic progress to be made
\cite{CGLref:halperin,CGLref:liu} in the form of
perturbation expansions.  Some numerical work has also
been performed on these systems \cite{CGLref:goldenfeld}.  For our system, the
numerical results are key as we cannot form perturbation
expansions due to the lack of an underlying Hamiltonian.
These numerical calculations
must begin with the determination of
the location of the defects.

To find the defects in our system, we examine the change in the phase
of $A$ as each plaquette (or square unit cell) on our lattice in real
space is traversed counterclockwise.  To do this, we examine the phase
$\varphi({\bf x},t) \equiv \tan^{-1} \left({{\rm Im}~A({\bf x},t)
\over {\rm Re}~A({\bf x}, t)} \right)$ as we go around a plaquette on
the lattice and sum up the phase differences between the four points
in the plaquette.  A change of $0$ signifies that the plaquette does
not contain a defect, while changes of $\pm 2 \pi$ reveal that a
defect exists in the plaquette.  We can actually speed up the defect
finding process a bit.  If all four sites in the plaquette have the
same sign of ${\rm Re}~[A]$ or of ${\rm Im}~[A]$, then the plaquette
cannot contain a defect.  More complicated situations can also be
handled, such as where one corner of the plaquette is of a different
sign from the other three corners for ${\rm Re}~[A]$, while the
opposite corner from the original one is of a different sign from the
other three corners for ${\rm Im}~[A]$.  In this case, the plaquette
cannot contain a defect.  These considerations avoid the costly
calculation of $\tan^{-1}$, and can be used for a large fraction of
the plaquettes in the lattice.

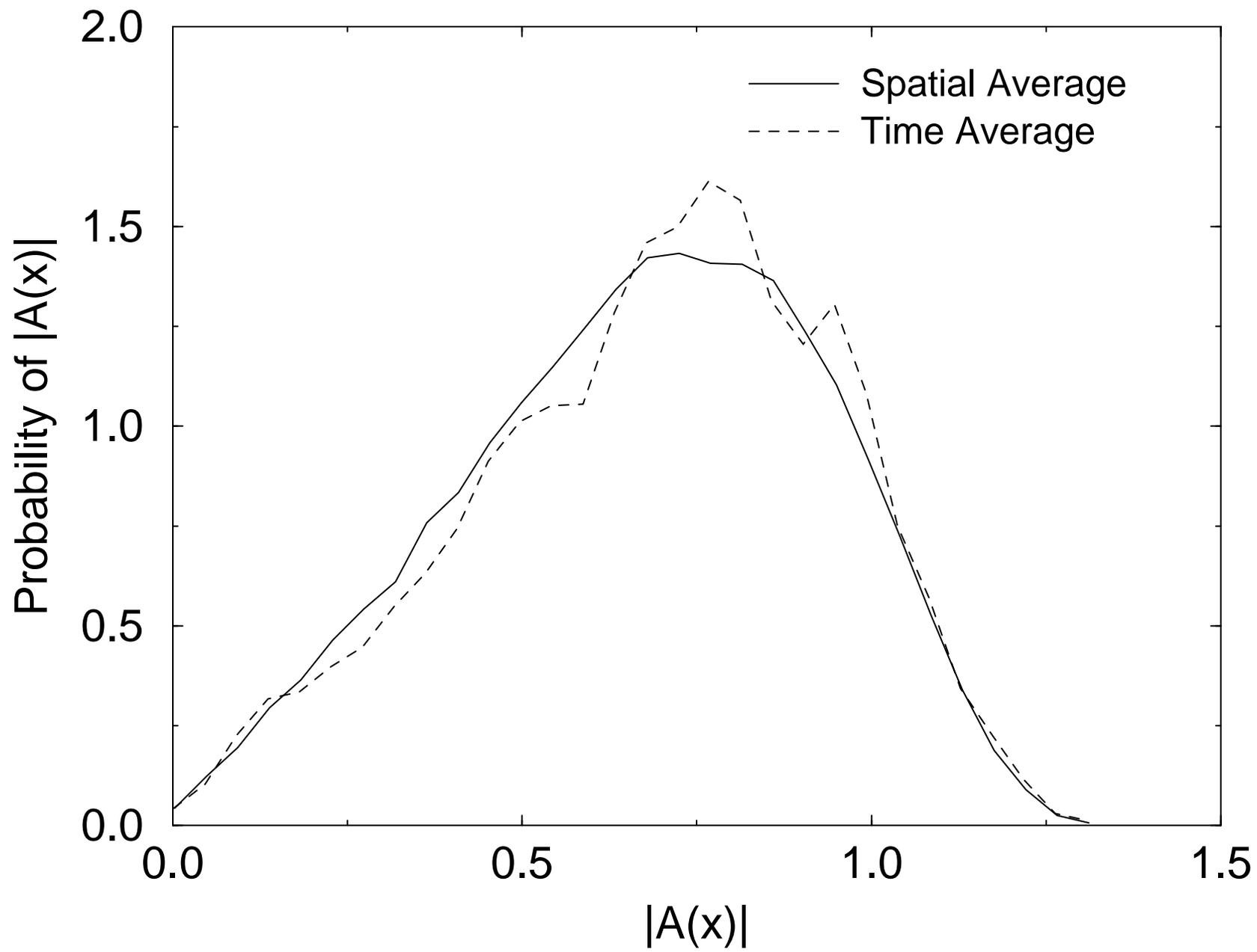
\begin{figure}
\caption[{Prob($n$)}]{
Prob($n$).  This was obtained from a time series of length $70,000$
time steps.
This is for a system of size $L_x = L_y = 240$ with $360$ Fourier
modes in each direction.  The parameter values are $c = 1.5$, $b_x =
-0.75$, and $b_y = -3.0$.  The dashed line is the exponential with
mean $\mu = 422.8$ and $\sigma^2 = 397$.
}
\label{CGLfig:probn}
\end{figure}

Once we have found the defects, we can calculate $n(t)$, the
total number of defects in the system, and $G_\rho({\rm \bf r})$.  To do
the averaging, we have run for $750,000$ time steps.
We only sample $G_\rho({\rm \bf r})$ and $n(t)$ every
$10$ time steps, because adjacent time steps are not statistically
independent.  We have calculated that
$\langle n(t) n(0) \rangle - \langle n \rangle^2 \sim {\rm e}^{-t/\tau}$,
with $\tau \sim 115$ time steps.
It has been predicted \cite{CGLref:defturbIII} that
the probability of finding a particular value of $n$ in the system is
given by $P(n) \sim {\rm e}^{-(n-\langle n \rangle)^2/2\langle n \rangle}$.
We have calculated the various moments of
our distribution $P(n)$, and we find $\langle n \rangle = 422.8 \pm 0.3$,
$\sigma^2 = 397 \pm 30$, as well as a skewness
of $0.014$ and a kurtosis of $-0.026$, which is in good agreement
with the predictions from reference \cite{CGLref:defturbIII}.
In Figure \ref{CGLfig:probn}, we present a plot of the results
for $P(n)$ together with the fitted exponential.  The agreement is
quite remarkable.

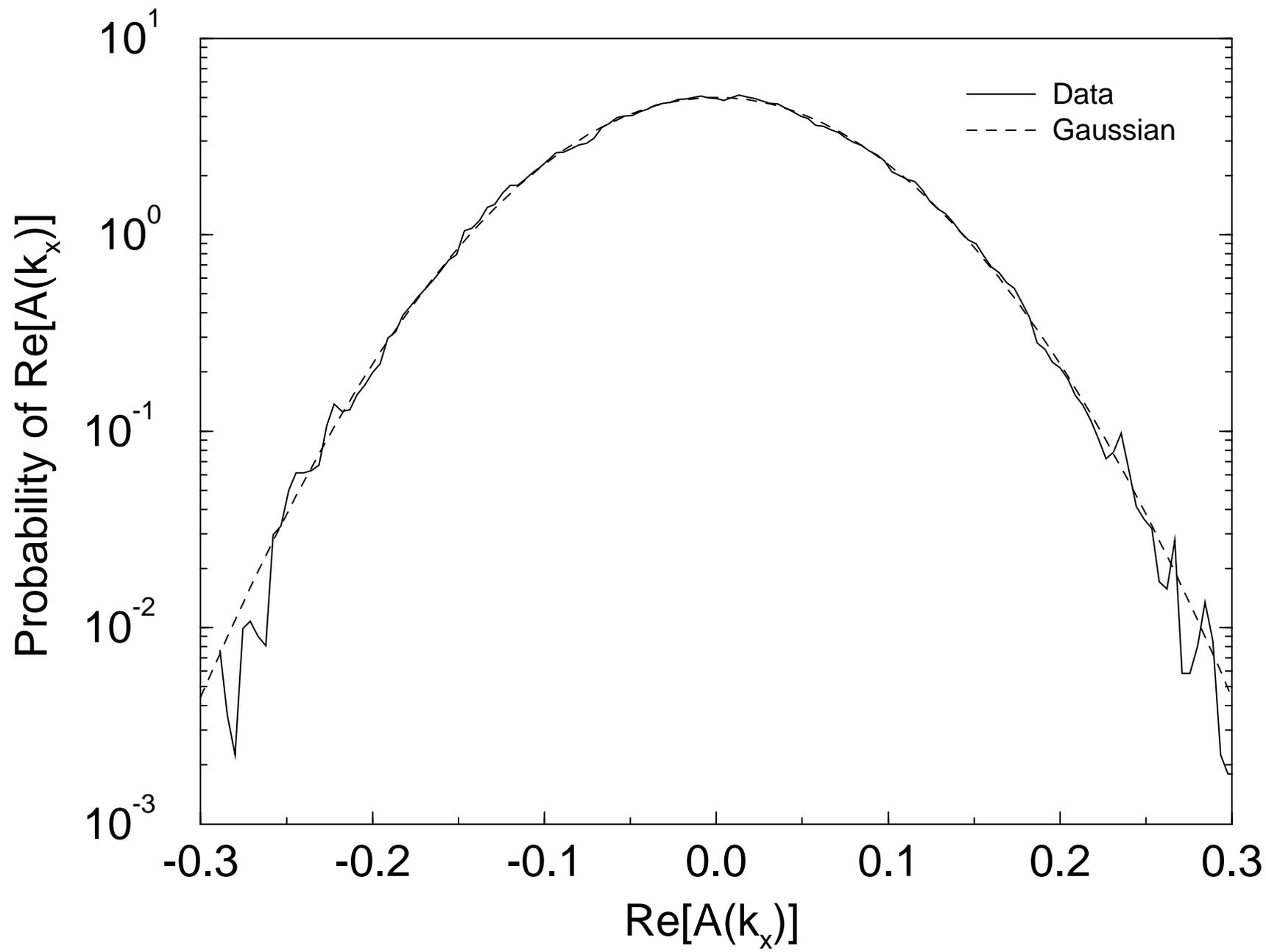
\begin{figure}
\caption[{$G_\rho({\rm \bf r})$ versus $r$}]
{
$G_\rho({\rm \bf r})$ versus $r$.  The solid line is for the ${\hat
x}$ direction, while the dashed line is for the ${\hat y}$ direction.
Also shown is a line for $G=0$.  Note that $G$ attains its asymptotic
limit of $0$ from different sides of this line.  The parameters are $c
= 1.5$, $b_x = -0.75$, and $b_y = -3.0$.
}
\label{CGLfig:grholin}
\end{figure}

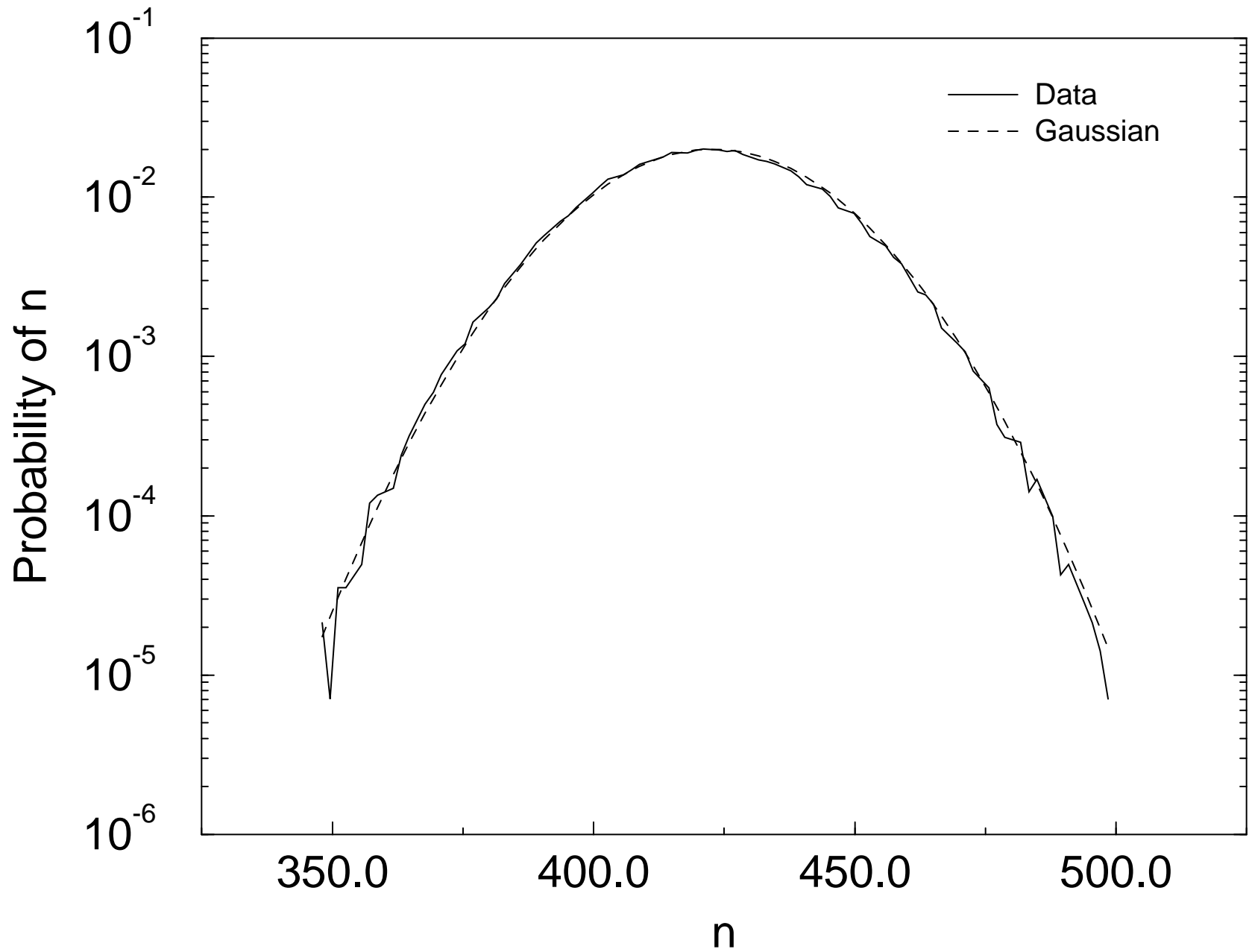
\begin{figure}
\caption[{$| G_\rho({\rm \bf r})| $ versus $r$ (linear--log plot)}]
{
$| G_\rho({\rm \bf r})| $ versus $r$ (linear--log plot).
The parameters are $c = 1.5$, $b_x = -0.75$, and $b_y = -3.0$.
}
\label{CGLfig:grholinlog}
\end{figure}

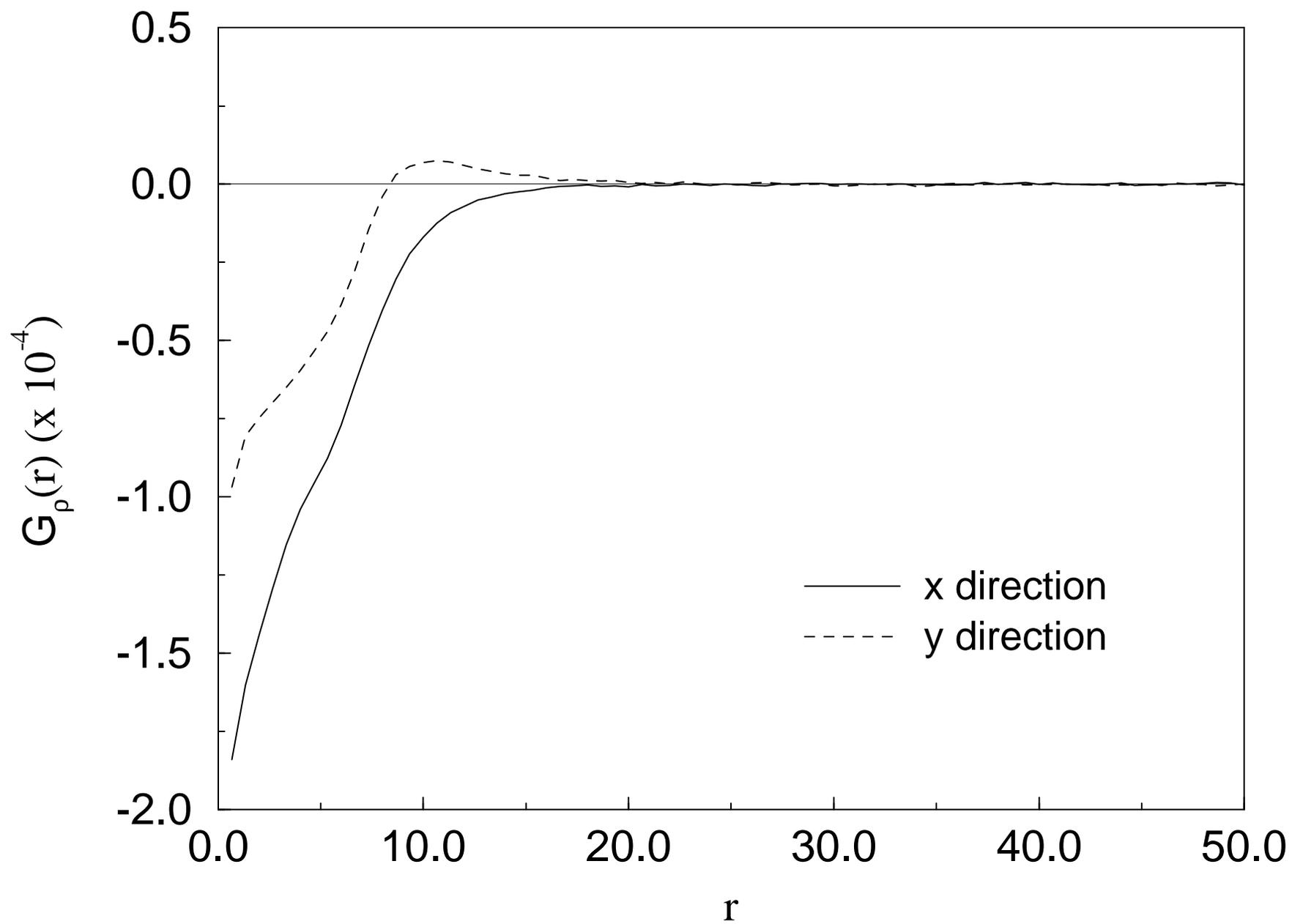
\begin{figure}
\caption[{Log--log plot of $| G_\rho({\rm \bf r}) |$ versus $r$}]
{
Log--log plot of $| G_\rho({\rm \bf r}) |$ versus $r$.
The solid line corresponds to the ${\hat x}$ direction and
the dashed line to the ${\hat y}$ direction.
Also shown is a line with slope that would correspond to
$|G_\rho(r)| \sim 1/r^2$.
The parameters are $c = 1.5$, $b_x = -0.75$, and $b_y = -3.0$.
}
\label{CGLfig:grhologlog}
\end{figure}

In Figure \ref{CGLfig:grholin} we present the results for
$G_\rho({\rm \bf r})$ with ${\rm \bf r}$ in both the $\hat x$ and $\hat y$
directions.  For both directions the typical nearest neighbor is of
the opposite sign: the charges are thus screened.  Similar behavior
for vortices in random wave fields has been observed
\cite{CGLref:zeros}.  In Figure \ref{CGLfig:grholinlog} we show a
linear--log plot of $|G_\rho(r)|$.  In Figure
\ref{CGLfig:grhologlog} we show log--log plots of $|G_\rho(r)|$.  We
also show lines that represent the slope $|G_\rho(r)|$ should have if
it decayed like $1/r^2$.  We note that at the right edge of the
figure, we have reached the point where our data is dominated by
statistical noise.  It is clear that neither direction shows the
expected $1/r^2$ decay.  Our results are at variance with the
predictions of generic scale invariance, as they must be. As was shown
in the last section, the theory is not applicable to systems which
have strong constraints placed on them due to the topological nature
of the order parameter.

The suggestion has been made \cite{CGLref:GrinPrivate}
that in order to see the $1/r^2$ decay the system needs to break
$x \rightarrow -x$ symmetry.  We systematically tested this idea by
adding various terms to the original equation (\ref{CGLeqn:cgl}).

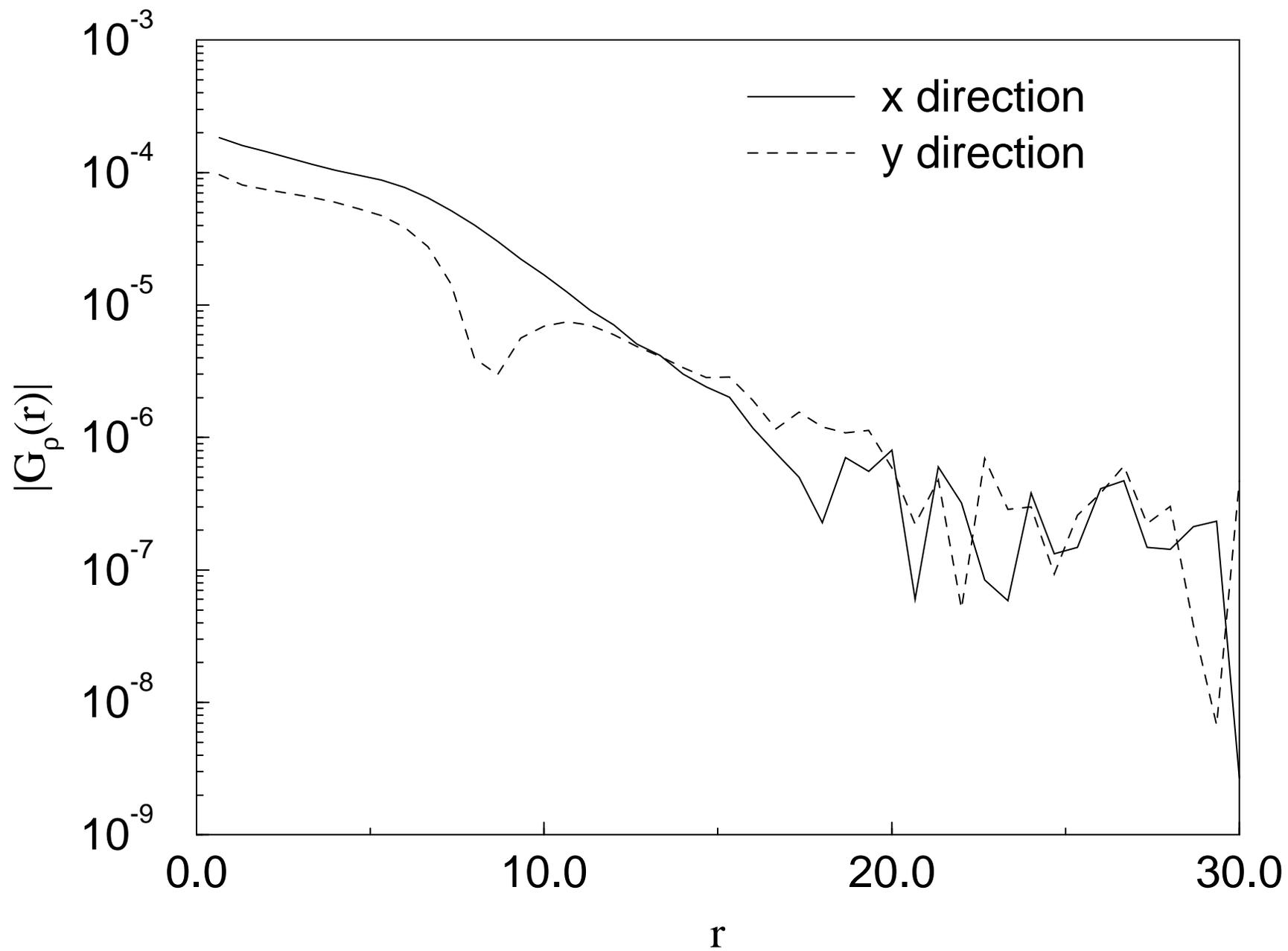
\begin{figure}
\caption[{Log--log plot of $| G_\rho({\rm \bf r}) |$ versus $r$ with
$d \partial_x A(x,t)$ added}]
{
Log--log plot of $| G_\rho({\rm \bf r}) |$ versus $r$
for the case where we have added a term $d \partial_x A(x,t)$ to
the complex Ginzburg--Landau equation.  Here we have $d = 0.1$.
The solid line corresponds to the ${\hat x}$ direction and
the dashed line to the ${\hat y}$ direction.
Also shown is a line with slope that would correspond to
$|G_\rho(r)| \sim 1/r^2$.
The parameters are $c = 1.5$, $b_x = -0.75$, and $b_y = -3.0$.
}
\label{CGLfig:grhologlog2}
\end{figure}

First, we added a term $d \partial_x A$ to the equation.
This term does break the $x \rightarrow -x$ symmetry.
The result for the correlation function $G_\rho(r)$ is given in
Figure \ref{CGLfig:grhologlog2}.  The results are
similar to the ones seen in Figure \ref{CGLfig:grhologlog} for
the results of the original equation.
Note that the term we have added here actually only induces a drift in
the entire pattern, and can be scaled away, so it is not surprising that
it does not show behavior that differs from the original equation.

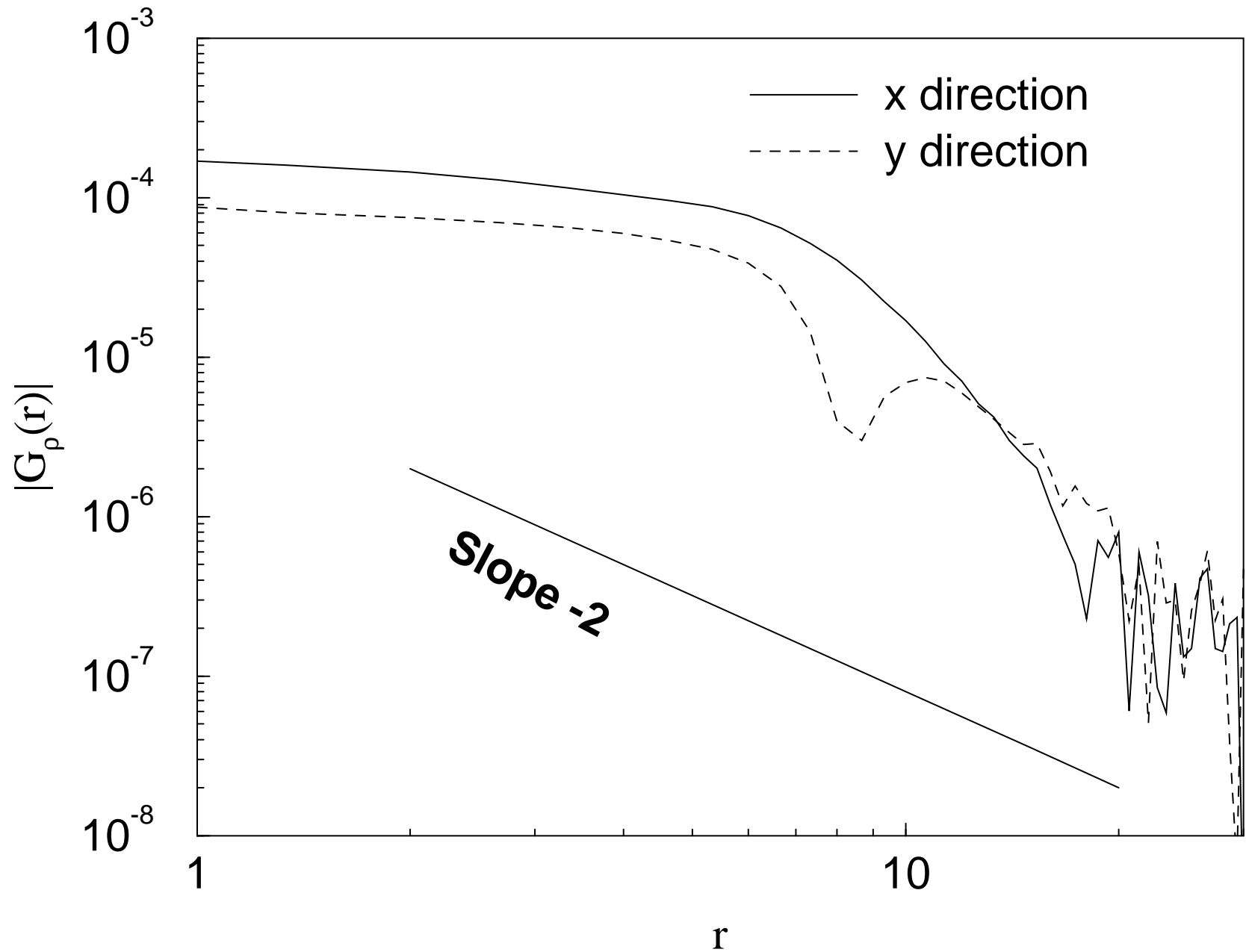
\begin{figure}
\caption[{Log--log plot of $| G_\rho({\rm \bf r}) |$ versus $r$ with
$d \partial_x^3 A(x,t)$ added}]
{
Log--log plot of $| G_\rho({\rm \bf r}) |$ versus $r$
for the case where we have added a term $d \partial_x^3 A(x,t)$ to
the complex Ginzburg--Landau equation.  Here we have $d = 0.1$.
The solid line corresponds to the ${\hat x}$ direction and
the dashed line to the ${\hat y}$ direction.
Also shown is a line with slope that would correspond to
$|G_\rho(r)| \sim 1/r^2$.
The parameters are $c = 1.5$, $b_x = -0.75$, and $b_y = -3.0$.
}
\label{CGLfig:grhologlog3}
\end{figure}

Second, we added the term $d \partial_x^3 A$ to the original equation.
The result for $G_\rho(r)$ is shown in Figure \ref{CGLfig:grhologlog3}.
This term also breaks the $x \rightarrow -x$ symmetry.
However, a term of this form is expected to show, according to the
ideas of generic scale invariance \cite{CGLref:GrinPrivate},
a decay that is actually faster than $1/r^2$, so our results are
not very informative.  We cannot distinguish between power laws
of $1/r^4$ and our results, so we must proceed to add
a different term to the original equation.

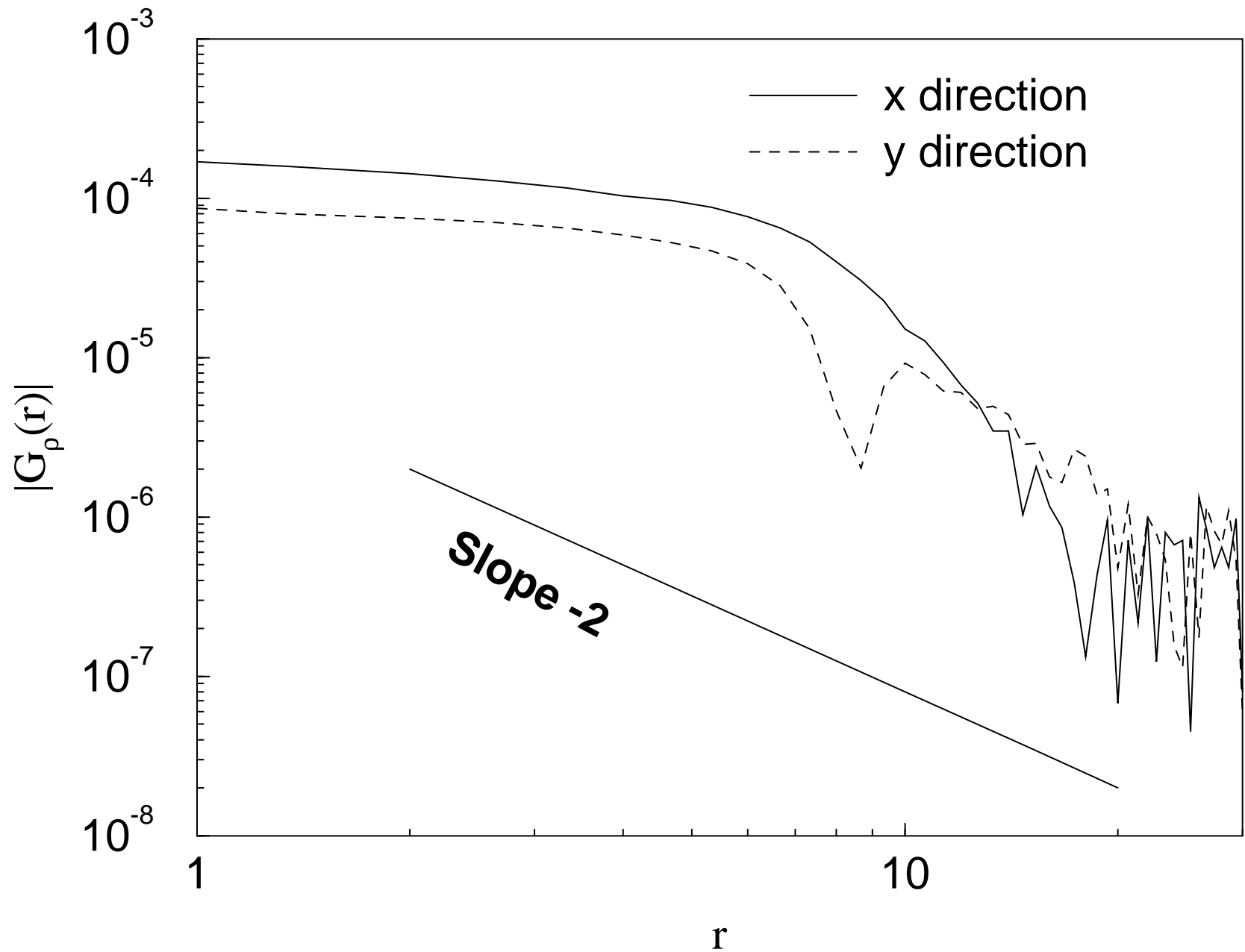
\begin{figure}
\caption[{Log--log plot of $| G_\rho({\rm \bf r}) |$ versus $r$ with
$d \partial_x |A|^2 A$ added}]
{
Log--log plot of $| G_\rho({\rm \bf r}) |$ versus $r$
for the case where we have added a term $d \partial_x |A(x,t)|^2 A(x,t)$ to
the complex Ginzburg--Landau equation.  Here we have $d = 1.0$.
The solid line corresponds to the ${\hat x}$ direction and
the dashed line to the ${\hat y}$ direction.
Also shown is a line with slope that would correspond to
$|G_\rho(r)| \sim 1/r^2$.
The parameters are $c = 1.5$, $b_x = -0.75$, and $b_y = -3.0$.
}
\label{CGLfig:grhologlog5}
\end{figure}

The third new term we added both breaks the $x \rightarrow -x$
symmetry and {\it is} expected to show a $1/r^2$ decay, according to the
predictions of generic scale invariance.
This term is $d \partial_x ( |A|^2 A)$.
The numerical results for $G_\rho(r)$
are shown in Figure \ref{CGLfig:grhologlog5}.
We see from this that the correlation function does not decay
like $1/r^2$.  This is not surprising in the light
of our results from Section \ref{CGLsec:constraints}, which apply
to any two--dimensional topological defect in a complex field.

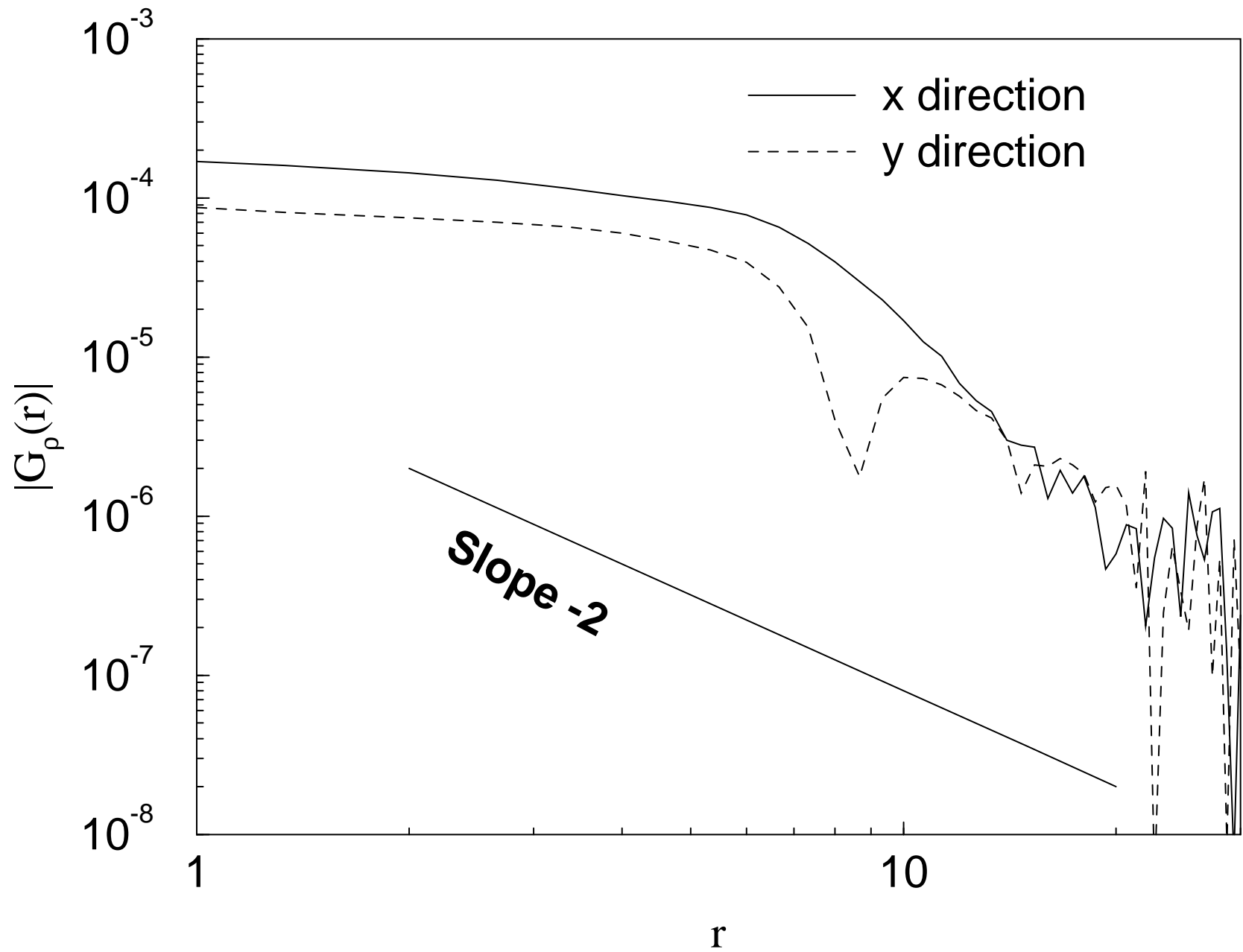
\begin{figure}
\caption[{Log--log plot of $| G_\rho({\rm \bf r}) |$ versus $r$ with
$d \partial_x^3 A(x,t)^2$ added}]
{
Log--log plot of $| G_\rho({\rm \bf r}) |$ versus $r$
for the case where we have added a term $d \partial_x^3 A(x,t)^2$ to
the complex Ginzburg--Landau equation.  Here we have $d = 1.0$.
The solid line corresponds to the ${\hat x}$ direction and
the dashed line to the ${\hat y}$ direction.
Also shown is a line with slope that would correspond to
$|G_\rho(r)| \sim 1/r^2$.
The parameters are $c = 1.5$, $b_x = -0.75$, and $b_y = -3.0$.
}
\label{CGLfig:grhologlog4}
\end{figure}

Finally, we experimented with breaking the
$A \rightarrow -A$ symmetry in the original equation by adding
the term $d \partial_x A^2$ to the original equation.  The results
for $G_\rho(r)$ are given in Figure \ref{CGLfig:grhologlog4}.
These results no longer shows the strong anisotropy seen in
the correlation function of the original unmodified equation.
With the broken symmetry, the decay
of the correlation function in the $\hat y$ direction
occurs smoothly to zero from the negative values near
$r = 0$ without crossing zero.

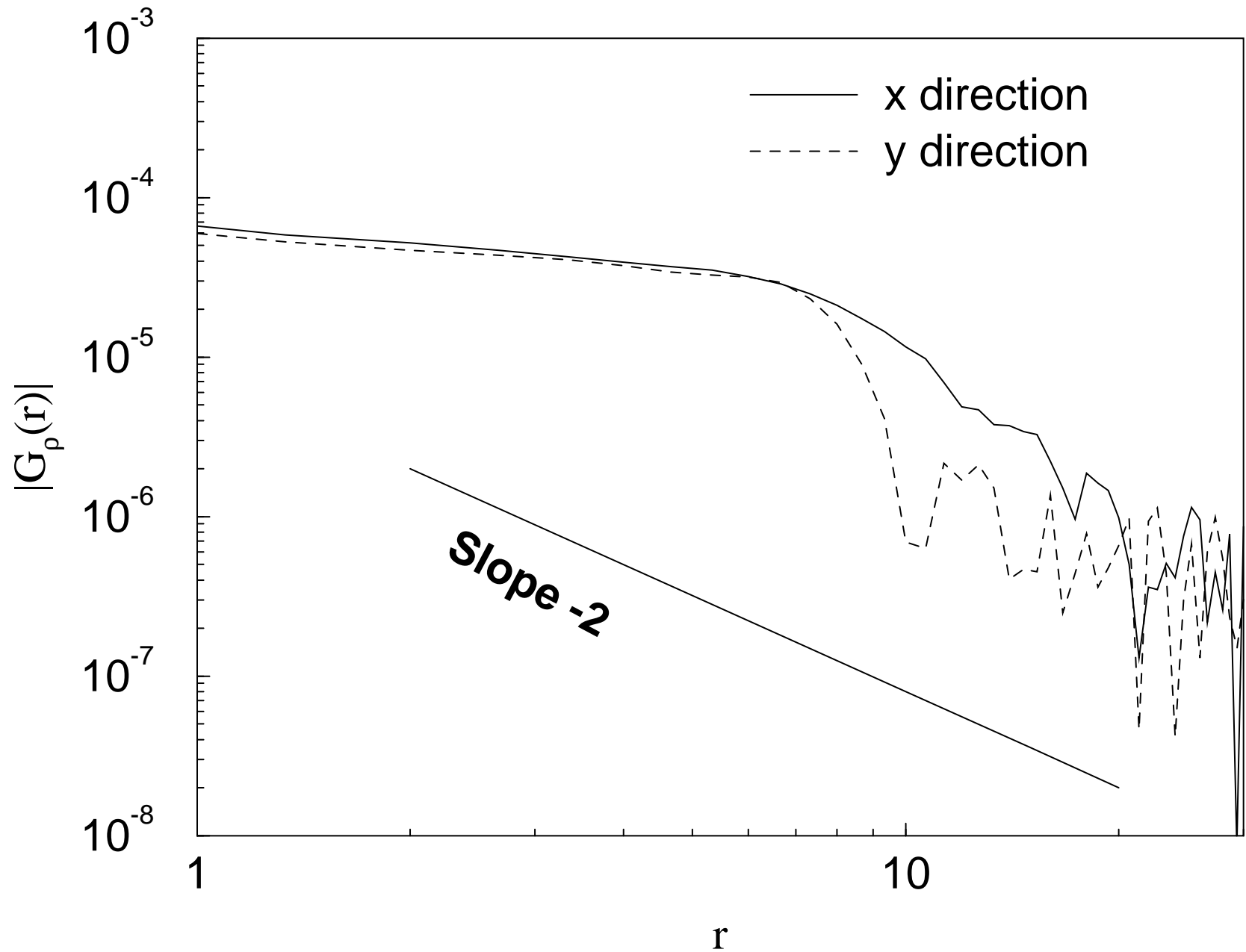
\begin{figure}
\caption[{Correlation function of the coarse--grained order parameter}]
{
Correlation function of the coarse--grained topological
order parameter.
The coarse graining scale is $\sigma = 10$.
The parameters are $c = 1.5$, $b_x = -0.75$, and $b_y = -3.0$.
}
\label{CGLfig:coarsecorr}
\end{figure}

We have also constructed a coarse--grained order parameter field
by defining
\begin{equation}
\rho_{cg}({\bf r}) \equiv \int {\rm d\bf s} \thinspace h({\bf s})
\rho({\bf r} - {\bf s})
\label{CGLeqn:cgdef}
\end{equation}
where $h({\bf s})$ is a coarse--graining weight function with
$\int {\rm d \bf s} \thinspace h({\bf s}) = 1$.  We will use the Gaussian
$h({\bf s}) = \frac{1}{2 \pi \sigma^2} {\rm e}^{-s^2/2 \sigma^2}$.
With this we can also calculate $G_{\rho_{cg}}(r)$.  The result
is shown in Figure \ref{CGLfig:coarsecorr}.
We can examine the leading behavior of this correlation function
by noting that
\begin{equation}
G_{\rho_{cg}}({\bf r}) = \langle \rho_{cg}({\bf r}) \rho_{cg}({\bf 0}) \rangle
	= \int \int {\rm d} {\bf s} {\rm d} {\bf s}' h({\bf s})
	h({\bf s}') \langle \rho({\bf r} - {\bf s}) \rho(-{\bf s}') \rangle.
\label{CGLeqn:cgcorrinit}
\end{equation}
This can be rewritten as
\begin{equation}
G_{\rho_{cg}}({\bf r}) = \int \int {\rm d} {\bf s} {\rm d} {\bf s}'
	\frac{1}{(2 \pi \sigma^2)^2} {\rm e}^{-\frac{1}{2 \sigma^2}
	(s^2 + s'^2)} G_\rho( {\bf r} - ({\bf s} - {\bf s}')).
\label{CGLeqn:cgcorrrelation}
\end{equation}
We now change variables $\mbox{\boldmath $\xi$} \equiv {\bf s} - {\bf s}'$ and
$\mbox{\boldmath $\eta$} \equiv {\bf s} + {\bf s}'$ to obtain
\begin{equation}
G_{\rho_{cg}}({\bf r}) = \frac{1}{(2 \pi \sigma^2)^2}
\int \int {\rm d} \mbox{\boldmath $\xi$}
{\rm d} \mbox{\boldmath $\eta$} {\rm e}^{-\eta^2/4 \sigma^2}
{\rm e}^{-\xi^2/4 \sigma^2} G_\rho({\bf r} - \mbox{\boldmath $\xi$}).
\label{CGLeqn:cgcorrvars}
\end{equation}
We can perform the $\eta$ integral to obtain the final result:
\begin{equation}
G_{\rho_{cg}}({\bf r}) = \frac{1}{\pi\sigma^2}
	\int {\rm d} \mbox{\boldmath $\xi$} {\rm e}^{-\xi^2/4\sigma^2}
	G_\rho({\bf r} - \mbox{\boldmath $\xi$}).
\label{CGLeqn:cgcorrfinal}
\end{equation}
Using this final equation, we can explain the Gaussian decay seen
in Figure \ref{CGLfig:coarsecorr}.  Typically, the
microscopic field $G_\rho({\bf r})$ will be given by
$ G_0 \delta({\bf r}) + g({\bf r})$.  Plugging this into
equation (\ref{CGLeqn:cgcorrfinal}) gives
\begin{equation}
G_{\rho_{cg}}({\bf r}) = \frac{G_0}{\pi \sigma^2}
	{\rm e}^{-r^2/4\sigma^2} + \frac{1}{\pi \sigma^2}
	\int {\rm d} \mbox{\boldmath $\xi$} {\rm e}^{-\xi^2/4\sigma^2}
	g({\bf r} - \mbox{\boldmath $\xi$}).
\label{CGLeqn:cggaussian}
\end{equation}
The behavior of the correlation function in Figure
\ref{CGLfig:coarsecorr} for small $r$ is approximately Gaussian, due
to the leading term in equation (\ref{CGLeqn:cggaussian}) as well
as corrections to the overall decay due to the second term in
equation (\ref{CGLeqn:cggaussian}).  For larger $r$ we see the domain
structure of the coarse grained field $\rho_{cg}$.  We do not see any
evidence of power--law decays in this coarse--grained correlation
function.

As a final comment on the numerical results presented in this section,
note that the typical spacing between defects in
our parameter regime is fairly large.  As a result,
we are not able to obtain adequate statistics for the correlation
functions for large $r$, and we might not
have reached the asymptotic regime
where generic scale invariance should apply.  Also, the
results for the noisier ($\hat y$) direction do not convincingly
rule out $1/r^2$ behavior. However, the proof we have provided
in Section \ref{CGLsec:constraints} shows analytically that the
topological nature of the defects precludes $1/r^2$ decay in the
correlation function.

\section{Conclusions}
\label{CGLsec:conclusions}

We have calculated a number of quantities of interest in the
complex Ginzburg--Landau equation.  These have included
both properties that depend on the amplitude field $A$ and
properties that depend upon the defect order parameter $\rho$.
The results for the field $A$ provide information about the
reliability of our numerics, as well as suggesting that the
Central Limit Theorem holds for $k$--space quantities.

The results for the order parameter field $\rho$ did not agree
with the predictions of generic scale invariance
\cite{CGLref:gsiI,CGLref:gsiII,CGLref:gsiIII}.  This places
a limit on the applicability of generic
scale invariance.  Conversely, the results
also also place a bound on the types
of coarse--grained, statistical theories that can be used to describe
spatiotemporally chaotic systems with topological defects.

\section{Acknowledgments}
\label{CLGsec:ack}

We would like to thank P.~C.~Hohenberg,
J.~F.~Marko, and M.~E.~J.~Newman
for helpful conversations.  This work was
partly funded by the Hertz Foundation~(BWR), the NSF under grant
DMR--91--18065~(BWR,~JPS), and the Alfred P. Sloan Foundation (EB).
We also thank the Cornell Materials Science Center for the use
of its computational resources.

\newpage

\newpage


%
%
%

%
%
\begin{figure}
\caption[{${\rm Re} [ \langle A^*({\bf r}) A(0) \rangle]$}]{
${\rm Re} [ \langle A^*({\bf r}) A(0) \rangle]$
averaged over $12$ different systems with $L_x = L_y = 240$.
The solid line is for ${\bf r} = r \hat x$ while the dashed line is for
${\bf r} = r \hat y$.
}
\label{CGLfig:AAcorr}
\end{figure}
%

%
%
%

%
%
%

%
%
%

%
%
\begin{figure}
\caption[{Prob(${\rm Re}~[A(k_x)]$)}]
{
Prob(${\rm Re}~[A(k_x)]$) averaged over $500,000$ time steps, with parameter
values $c = 1.5$, $b_x = -1.5$, and $b_y = -1.5$.  Also shown in the
dashed line is a Gaussian with mean $0$ and standard deviation $0.08$ for
comparison.  We have chosen a value of $k = \frac{2 \pi (2)}{60}$.  The
system has size $L_x = L_y = 60$ with $90$ Fourier modes in each direction.
}
\label{CGLfig:probAk}
\end{figure}
%

%
%
%

%
%
%

%
%
%

%
%
%

%
%
%

%
%
%

%
%
%

%
%
%

%
%

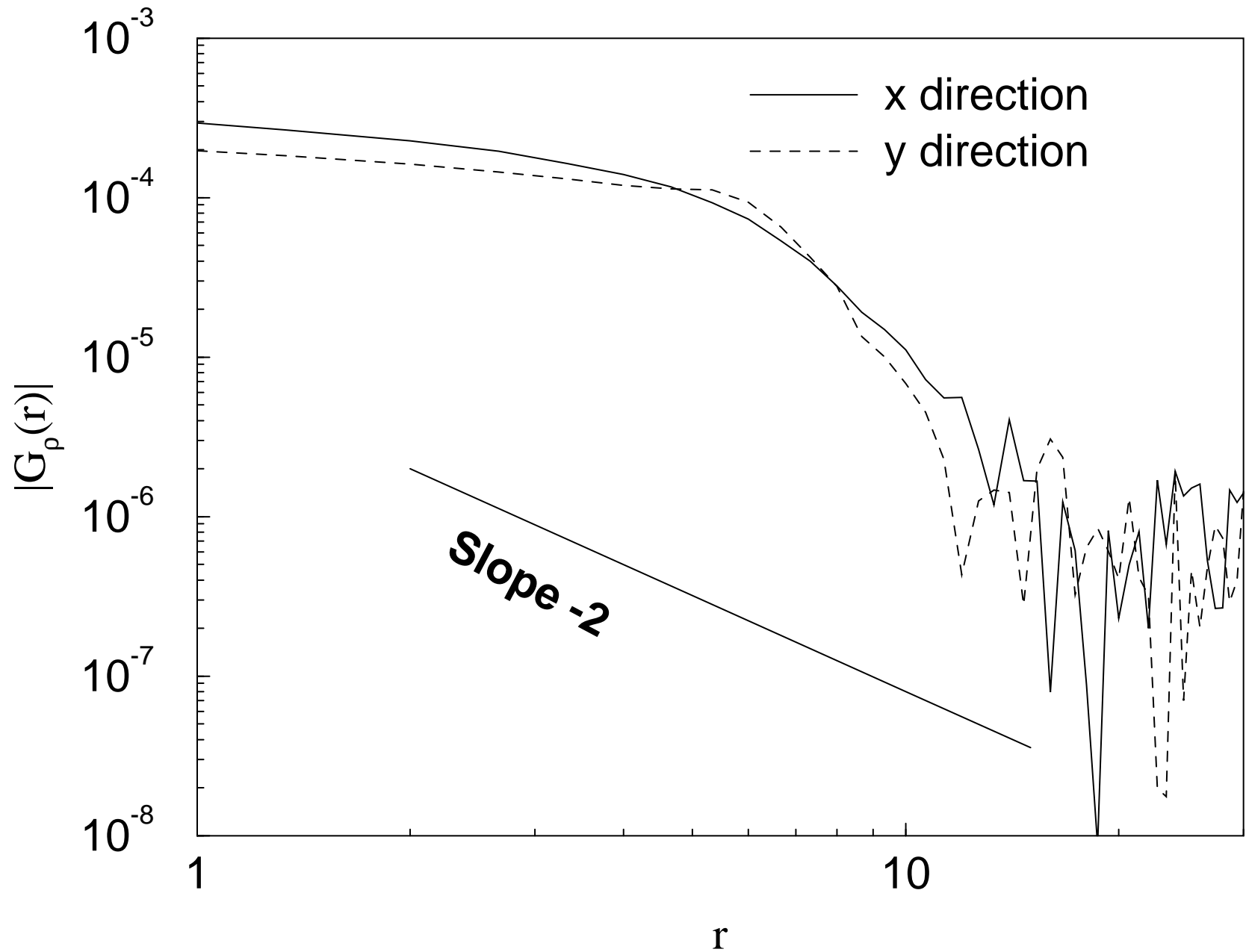
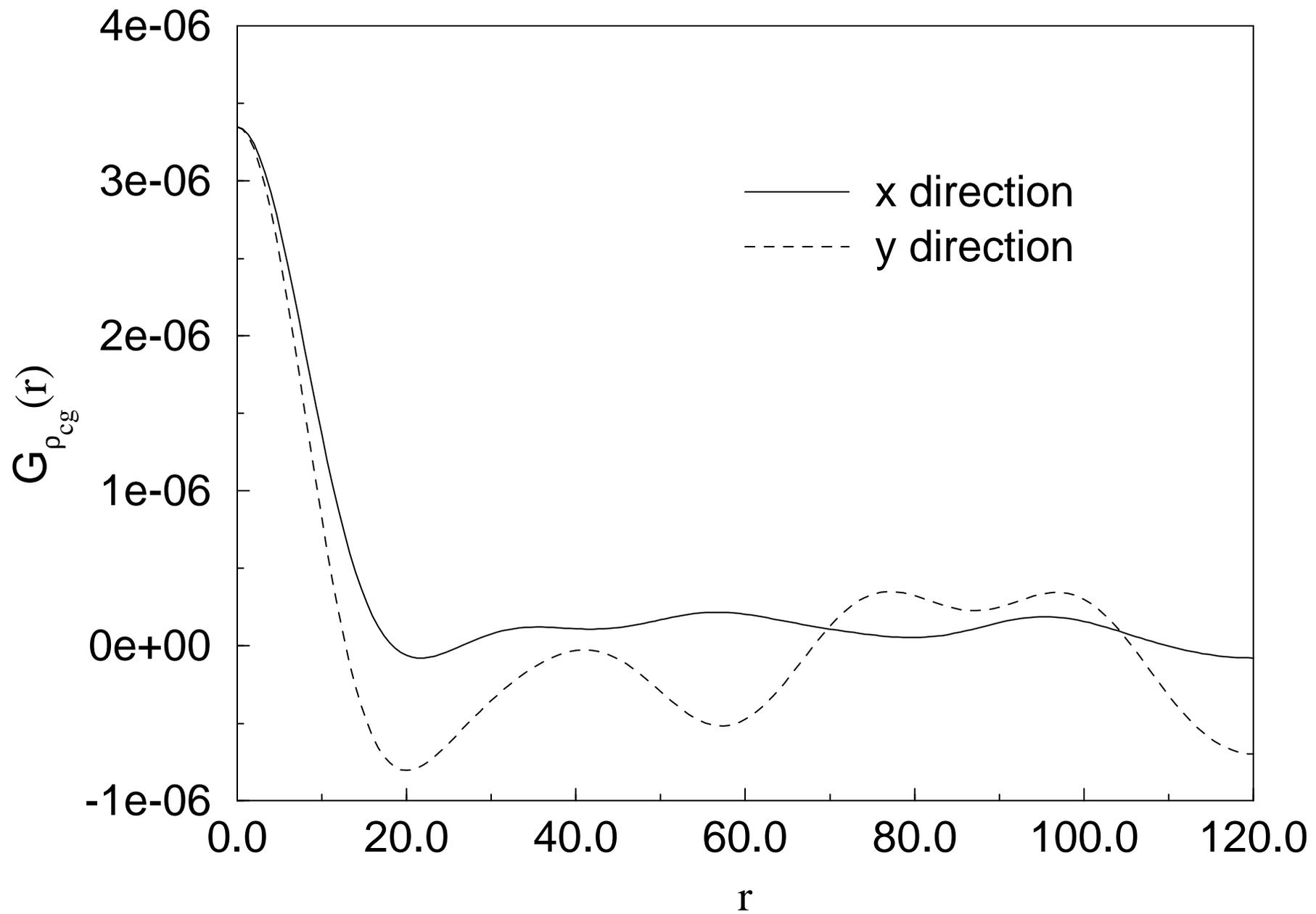
\end{document}